\documentclass{elsarticle}
\usepackage[utf8]{inputenc}
\usepackage[toc,page]{appendix}
\usepackage{amsmath}
\usepackage{amsfonts}
\usepackage{tikz}
\usetikzlibrary{patterns}
\usepackage{subcaption}
\usepackage[a4paper, total={7in, 9in}]{geometry}
\usepackage{indentfirst}
\usepackage{amsthm}
\usepackage{natbib}
\usepackage{graphicx}
\usepackage{hyperref}
\usepackage{cleveref}
\usepackage{float}
\usepackage{xcolor}
\usepackage{multicol}
\usepackage{nomencl}
\usepackage{bm}

\makenomenclature

\usepackage{etoolbox}
\renewcommand\nomgroup[1]{%
  \item[\bfseries
  \ifstrequal{#1}{A}{Symbols}{%
  \ifstrequal{#1}{B}{Operators}{}}%
]}

\usepackage{upgreek}

\Crefname{equation}{Eq.}{Eqs.}
\Crefname{figure}{Fig.}{Figs.}
\Crefname{table}{Tab.}{Tabs.}
\Crefname{appsec}{Appendix}{Appendices}
\crefname{appsec}{appendix}{appendices}

\makeatletter
\def\ps@pprintTitle{%
   \let\@oddhead\@empty
   \let\@evenhead\@empty
   \def\@oddfoot{\reset@font\hfil\thepage\hfil}
   \let\@evenfoot\@oddfoot
}
\makeatother

\begin{document}

\begin{abstract}
A homogenization approach is proposed for the treatment of porous wall boundary conditions in the computation of compressible viscous flows. Like any other homogenization approach,
it eliminates the need for pore-resolved fluid meshes and therefore enables practical flow simulations in computational fluid domains with porous wall boundaries. Unlike alternative approaches
however, it does not require prescribing a mass flow rate and does not introduce in the computational model a heuristic discharge coefficient. Instead, it models the inviscid flux through a porous 
wall surrounded by the flow as a weighted average of the inviscid flux at an impermeable surface and that through pores. It also introduces a body force term in the governing equations to account 
for friction loss along the pore boundaries. The source term depends on the thickness of the porous wall and the concept of an equivalent single pore.  The feasibility of the
latter concept is demonstrated using low-speed permeability test data for the fabric of the Mars Science Laboratory parachute canopy. The overall homogenization approach is illustrated with a 
series of supersonic flow computations through the same fabric and verified using supersonic, pore-resolved numerical simulations. 

\end{abstract}

\begin{frontmatter}  
\title{Homogenized Flux-Body Force Treatment of Compressible Viscous Porous Wall Boundary Conditions}
\author[rvt1]{Daniel~Z.~Huang}
\ead{zhengyuh@stanford.edu}

\author[rvt2]{Man Long Wong}
\ead{wongml@stanford.edu}

\author[rvt1,rvt2,rvt3]{Sanjiva~K.~Lele}
\ead{lele@stanford.edu}

\author[rvt1,rvt2,rvt3]{Charbel~Farhat}
\ead{cfarhat@stanford.edu}

\address[rvt1]{Institute for Computational and Mathematical Engineering,
               Stanford University, Stanford, CA, 94305}
\address[rvt2]{Department of Aeronautics and Astronautics, Stanford University, Stanford, CA, 94305}
\address[rvt3]{Department of Mechanical Engineering, Stanford University, Stanford, CA, 94305}
\end{frontmatter}

\nomenclature[A]{$\rho$}{fluid density}
\nomenclature[A]{$\bm{W}$}{vector of conservative fluid state variables}
\nomenclature[A]{$\bm{V}$}{vector of primitive fluid state variables}
\nomenclature[A]{$t$}{time}
\nomenclature[A]{$\mathcal{F}$}{physical inviscid flux tensor}
\nomenclature[A]{$\mathcal{G}$}{physical viscous flux tensor}
\nomenclature[A]{$\bm{v}$}{fluid velocity vector}
\nomenclature[A]{$u$}{$x$-component of the fluid velocity vector}
\nomenclature[A]{$v$}{$y$-component of the fluid velocity vector}
\nomenclature[A]{$w$}{$z$-component of the fluid velocity vector}
\nomenclature[A]{$p$}{fluid pressure}
\nomenclature[A]{$E$}{fluid total energy per unit volume}
\nomenclature[A]{$e$}{fluid specific internal energy}
\nomenclature[A]{$T$}{fluid temperature}
\nomenclature[A]{${\uptau}$}{fluid viscous stress tensor}
\nomenclature[A]{$\bm{q}$}{heat flux vector} 
\nomenclature[A]{$\mu$}{dynamic shear viscosity}
\nomenclature[A]{$\kappa$}{fluid thermal conductivity}
\nomenclature[A]{$R$}{gas constant} 
\nomenclature[A]{$n$}{unit outward normal to a surface} 
\nomenclature[A]{$\gamma$}{specific heat ratio}
\nomenclature[A]{$\alpha$}{void fraction or porosity}
\nomenclature[A]{$\delta_x$}{mesh spacing in $x$-direction}
\nomenclature[A]{$\delta_y$}{mesh spacing in $y$-direction}
\nomenclature[A]{$\delta_z$}{mesh spacing in $z$-direction}
\nomenclature[A]{$C$}{velocity parameter of a Hagen-Poiseuille or similar type of flow}
\nomenclature[A]{$r$}{pore size parameter}
\nomenclature[A]{$\Omega$}{surface representation of a porous wall of finite thickness}
\nomenclature[A]{$\Omega_{\textrm{vol}}$}{volume representation of a porous wall of finite thickness}
\nomenclature[A]{$\partial\Omega_{\textrm{vol}}$}{wet boundary surface of $\Omega_{\textrm{vol}}$}
\nomenclature[A]{$|\Omega|$}{porous wall area}
\nomenclature[A]{$\Omega^{\textrm{pore}}$}{part of $\Omega$ occupied by the pores}
\nomenclature[A]{$\Omega_{\textrm{vol}}^{\textrm{pore}}$}{part of $\Omega_{\textrm{vol}}$ occupied by the pores}
\nomenclature[A]{$\partial\Omega_{\textrm{vol}}^{\textrm{pore}}$}{boundary surface of $\Omega_{\textrm{vol}}^{\textrm{pore}}$}
\nomenclature[A]{$|\Omega^{\textrm{pore}}|$}{pore area}
\nomenclature[A]{$\eta$}{porous wall thickness}
\nomenclature[A]{$B$}{computational fluid domain}
\nomenclature[A]{$\bm{X}$}{position vector of a point in $B$}
\nomenclature[A]{$B^e$}{primal element of the discretization of $B$}
\nomenclature[A]{$|B^e|$}{volume of $B^e$}
\nomenclature[A]{${\partial B}_{\infty}$}{far-field boundary of $B$}
\nomenclature[A]{${\mathcal C}_i$}{control volume (dual cell) attached to a grid point $i$}
\nomenclature[A]{$|{\mathcal C}_i|$}{volume of ${\mathcal C}_i$}
\nomenclature[A]{${\mathcal K}(i)$}{set of grid points connected to the grid point $i$}
\nomenclature[A]{$\partial {\mathcal C}_i$}{boundary surface of ${\mathcal C}_i$}
\nomenclature[A]{$\partial {\mathcal C}_{ij}$}{boundary surface (cell facet) shared by ${\mathcal C}_i$ and ${\mathcal C}_j$}
\nomenclature[A]{$|\partial{\mathcal C}_{ij}|$}{area of $\partial{\mathcal C}_{ij}$}
\nomenclature[A]{$\psi_i$}{piecewise-linear tent test function associated with the grid point $i$}
\nomenclature[A]{$\bm{F}$}{numerical inviscid flux vector function in the $x$-direction}
\nomenclature[A]{$\bm{G}$}{numerical viscous flux vector in the $x$-direction}
\nomenclature[A]{$\bm{S}$}{source term or body force vector}
\nomenclature[A]{$x_0$}{position of a porous wall along the $x$-direction}
\nomenclature[A]{{$\eta_f$}}{thickness correction factor}
\nomenclature[A]{$V_s$}{shock speed}
\nomenclature[A]{$Re$}{Reynolds number}
\nomenclature[A]{$d$}{diameter of an equivalent circular pore}
\nomenclature[A]{$K$}{constant of proportionality between the fluid mass flow rate and pressure difference}
\nomenclature[A]{$F_D$}{drag force}
\nomenclature[B]{$\Delta \phi$}{difference across the porous wall between the values of the variable $\phi$}
\nomenclature[B]{$D(\phi)$}{Gaussian function of variable $\phi$}
\nomenclature[B]{$\overline{\phi}$}{homogenized variable $\phi$}
\nomenclature[B]{$\Gamma$}{interface between two fluid subdomains}

\begin{multicols}{2}
\printnomenclature
\end{multicols}

\section{Introduction}

Porous walls and membranes appear in a wide range of scientific and engineering applications. These include, to name only a few, 
filtration processes \cite{nassehi1998modelling, griffiths2013control, ling2016dispersion} to separate particulates or molecules from a bulk fluid, 
the cooling of gas turbines via multiperforated plates \cite{apte2001unsteady, mendez2008large, mendez2008adiabatic, mendez2007large},
light weight canopies for low density supersonic decelerators \cite{kim20062, karagiozis2011computational, gao2016numerical, huang2018simulation, huang2020modeling}, 
and windbreaks or shelterbelts \cite{judd1996wind, patton1998large, wang2001shelterbelts} to reduce the relative flow speed and maintain stability. 
For most of these large-scale engineering applications, pore-resolved CFD computations remain unpractical, if not unaffordable. Therefore, the 
high-precision and yet computationally efficient treatment of porous wall boundary conditions in CFD computations is crucial for enabling flow
simulations that further the understanding of the effects of porous materials and can assist in the development of membrane-related processes and technologies. 

The modeling of fluid flow through a porous wall or membrane has been inspired by a variety of investigations of flows in porous media. In \cite{kim20062, tutt2010development,
takizawa2012fluid, gao2016numerical, takizawa2017porosity}, a modified Darcy's law was coupled with Ergun's equation to construct an internal pressure jump condition for the incompressible or 
low-Mach Navier-Stokes equations and account for the material porosity of a canopy in the simulation of parachute fluid-structure interactions. 
In \cite{wilson1985numerical} and \cite{wang2001shelterbelts}, the flow associated with a windbreak similar to a plant canopy was modeled by adding a local, momentum 
extraction, source term near the windbreak. The pressure-loss coefficient in the source term was calibrated using the pressure drop across the windbreak, as well as the flow momentum measured 
in related wind tunnel experiments \cite{judd1996wind}. In \cite{nassehi1998modelling} and \cite{griffiths2013control}, the analysis of the cross-flow filtration was performed by numerically
simulating the transport of solutes or particles in a tube with porous walls using the Navier-Stokes equations and applying the slip wall boundary condition, as advocated in \cite{beavers1967boundary}. 
Notably, all of the aforementioned modeling approaches are limited to the incompressible flow regime. 

Only a few approaches for modeling the effects of permeability on high Mach number compressible flows have been documented so far. Those approaches have combined Darcy's law with the continuity 
equation \cite{muskat1934flow}, or derived compressible variants of Darcy's law from the continuity equation and the force balance equation between drag and the pressure drop across an element of a
porous material \cite{shepherd1988transient, schmidt2014compressible}. For flow computations associated with film-cooling in gas turbines operating at high-speed, the following works are noteworthy.
In \cite{apte2001unsteady}, the treatment of the porous wall boundary condition was performed using a mass inflow condition with white noise in the mass flow rate. In \cite{mendez2008adiabatic},
a homogeneous model was proposed for a multiperforated plate where the homogenized mass flux is related to a constant discharge coefficient estimated from pore-resolved numerical simulations. 

Here, an alternative homogenization approach is proposed for the treatment of porous wall boundary conditions in the numerical simulation of compressible viscous flows through permeable surfaces. 
Unlike the aforementioned approaches, it does not require prescribing a mass flow rate nor introduces an empirical discharge coefficient in the computational model. Instead, it models the inviscid flux
through a porous wall as a weighted average of the inviscid flux at an impermeable surface and that through pores. To account for the loss of friction along the pore boundaries, it introduces a 
body force term in the governing equations that depends on the thickness of the porous wall and the concept of an ``equivalent pore'' -- that is, a {\it single} pore whose geometry and 
dimensions are such that when inserted at the center of a unit cell, the void fraction of the cell matches that of the fabric of interest. The feasibility of such a concept is demonstrated using 
low-speed permeability experimental data for the fabric of the Mars Science Laboratory parachute canopy. The overall homogenization approach can be implemented in any finite difference,
finite volume, or finite element semi-discretization scheme. It is illustrated with a series of supersonic porous 
flow problems with Martian atmosphere free-stream conditions and verified using pore-resolved numerical simulations.

The remainder of the paper is organized as follows. First, the governing equations are specified in \Cref{sec: gov}. Next, the proposed homogenization approach is presented in \Cref{sec: model} and
its verification is performed in \Cref{sec: app}. Finally, conclusions are offered in \Cref{sec: conclusion}.

\section{Governing equations}\label{sec: gov}

Throughout this paper, flows are assumed to be compressible as well as viscous and are governed by the Navier-Stokes equations. These can be written in conservation form as follows
\begin{equation} \label{eq: ns}
\dfrac{\partial \bm{W}}{\partial t} +  \nabla\cdot \mathcal{F}(\bm{W}) =  \nabla\cdot \mathcal{G}(\bm{W})\,,
\end{equation}
where $\bm{W}$ denotes the vector of the conservative variables describing the fluid state, $t$ denotes time, and $\mathcal{F}(\bm{W})$ and $\mathcal{G}(\bm{W})$ are the inviscid and viscous flux tensor 
functions of the conservative variables, respectively. Specifically,
\begin{equation}
\bm{W} = 
\begin{pmatrix}
\rho \\[4mm]
\rho \bm{v} \\[4mm]
E
\end{pmatrix},
\quad 
\mathcal{F}(\bm{W}) = 
\begin{pmatrix}
\rho \bm{v} \\[2mm]
\rho \bm{v} \otimes \bm{v} + p\mathcal{I} \\[2.5mm]
\bigl(E + p \bigr)\bm{v}
\end{pmatrix},
\quad  \mathrm{and} \quad
\mathcal{G}(\bm{W}) =
\begin{pmatrix}
0 \\[2.5mm]
{\uptau} \\[1.5mm]
{\uptau}\boldsymbol \cdot \bm{v} - \bm{q}
\end{pmatrix},
\end{equation}
where $\rho$, $\bm{v}$, and $E$ denote the density, velocity, and total energy per unit volume of the fluid, respectively, and $\otimes$ denotes the Hadamard product. The fluid velocity and total
energy per unit volume can be written as
\begin{equation}
    \bm{v} = (u, v, w)^{T} \quad \textrm{and} \quad E = \rho e + \dfrac{1}{2}\rho (u^2 + v^2 + w^2)\,,
\end{equation}
where the superscript $T$ designates the transpose and $e$ denotes the specific (i.e., per unit of mass) internal energy. In the physical inviscid flux tensor $\mathcal{F}$, $p$ is the static pressure and $\mathcal{I} \in \mathbb{R}^3$ is the identity matrix. In the viscous flux tensor $\mathcal{G}$, ${\uptau}$ and $\bm{q}$ denote the viscous stress tensor and heat flux vector, respectively, and are defined as follows
\begin{equation}
{\uptau} = \mu( \nabla^T \bm{v} +  \nabla \bm{v}) - \dfrac{2}{3}\mu(\nabla\boldsymbol{\cdot} \bm{v})\mathcal{I}
\quad \mathrm{and} \quad
\bm{q} = -\kappa \nabla T,
\end{equation}
where $\mu$ is the dynamic shear viscosity of the fluid, $\kappa$ is its thermal conductivity, and $T$ is its temperature.

For simplicity, but without any loss of generality, the system of equations (\ref{eq: ns}) is closed by assuming that the gas is ideal, calorically perfect, and therefore governed by
\begin{equation}
    p = \rho R T \quad \mathrm{and} \quad e = \frac{R}{\gamma - 1} T,
\end{equation}
where $R$ and $\gamma$ are the gas constant and specific heat ratio, respectively.

\section{Homogenized flux-body force approach}
\label{sec: model}

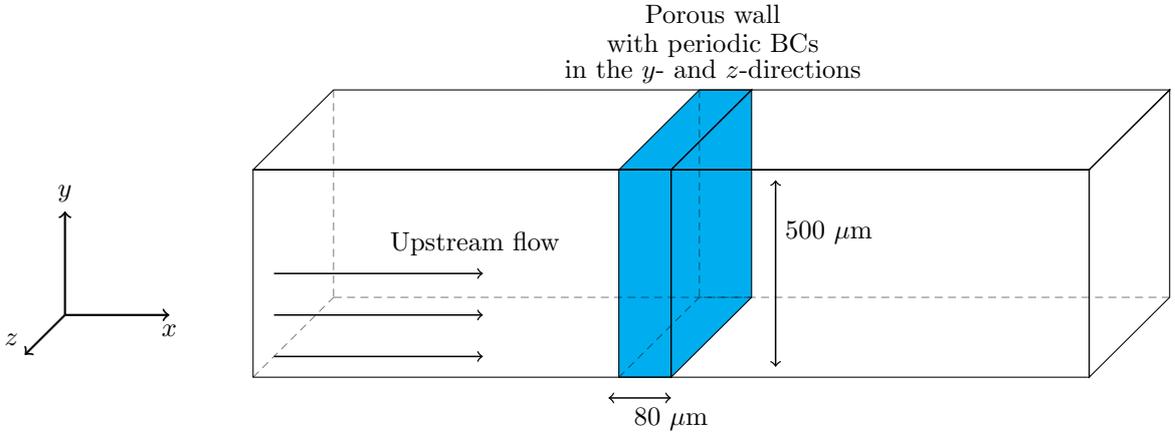
\begin{figure}[!ht]
\centering
    \centering
    \begin{tikzpicture}[scale=2.75]
    \draw[thick,->] (-0.9, 0.3, 0) -- (-0.4, 0.3, 0) node[anchor=north]{$x$}; 
    \draw[thick,->] (-0.9, 0.3, 0) -- (-0.9, 0.8, 0) node[anchor=south]{$y$};
    \draw[thick,->] (-0.9, 0.3, 0) -- (-0.9, 0.3, 0.5) node[xshift=-5][anchor=south]{$z$};
    
    \draw[thick,->,line width=0.2mm,](0.1, 0.5) -- (1.1, 0.5) node[xshift=-3,yshift=3][anchor=south][scale=1.0] {Upstream flow};
    \draw[thick,->,line width=0.2mm,](0.1, 0.3) -- (1.1, 0.3);
    \draw[thick,->,line width=0.2mm,](0.1, 0.1) -- (1.1, 0.1);
    
    \draw node at (2.2, 1.75) {Porous wall}; 
    \draw node at (2.2, 1.60) {with periodic BCs};
    \draw node at (2.2, 1.475) {in the $y$- and $z$-directions};
    
    \draw[thick,<->,line width=0.2mm,](2 + 0.5, 0.05) -- (2 + 0.5, 0.95) node[yshift=-12][anchor=north west][scale=1.0] {$500\ \mathrm{\mu m}$};
    \draw[thick,<->,line width=0.2mm,](2 - 0.3, -0.1) -- (2 , -0.1) node[anchor=north][scale=1.0] {$80\ \mathrm{\mu m}$};
    
    \pgfmathsetmacro{\cubex}{0.25}
    \pgfmathsetmacro{\cubey}{1}
    \pgfmathsetmacro{\cubez}{1}
    \draw [draw=black, every edge/.append style={draw=black, densely dashed, opacity=.5}, fill=cyan]
    (2,1,0) coordinate (o) -- ++(-\cubex,0,0) coordinate (a) -- ++(0,-\cubey,0) coordinate (b) edge coordinate [pos=1] (g) ++(0,0,-\cubez)  -- ++(\cubex,0,0) coordinate (c) -- cycle
    (o) -- ++(0,0,-\cubez) coordinate (d) -- ++(0,-\cubey,0) coordinate (e) edge (g) -- (c) -- cycle
    (o) -- (a) -- ++(0,0,-\cubez) coordinate (f) edge (g) -- (d) -- cycle;

    \pgfmathsetmacro{\cubex}{4}
    \pgfmathsetmacro{\cubey}{1}
    \pgfmathsetmacro{\cubez}{1}
    \draw [draw=black, every edge/.append style={draw=black, densely dashed, opacity=.5}]
    (4,1,0) coordinate (o) -- ++(-\cubex,0,0) coordinate (a) -- ++(0,-\cubey,0) coordinate (b) edge coordinate [pos=1] (g) ++(0,0,-\cubez)  -- ++(\cubex,0,0) coordinate (c) -- cycle
    (o) -- ++(0,0,-\cubez) coordinate (d) -- ++(0,-\cubey,0) coordinate (e) edge (g) -- (c) -- cycle
    (o) -- (a) -- ++(0,0,-\cubez) coordinate (f) edge (g) -- (d) -- cycle; 
    \end{tikzpicture}
	\caption{Porous flow model problem: porous wall, computational fluid domain, and periodic boundary conditions.}
\label{fig: computational domain}
\end{figure}

Consider as a backdrop for the homogenization approach described herein the three-dimensional, porous flow model problem graphically depicted in \Cref{fig: computational domain}. The
computational fluid domain is a three-dimensional, orthogonal box with a square cross section (the generalization to an arbitrarily-shaped computational fluid domain is straightforward). 
It contains at its center a porous wall of a finite thickness (shown in blue color) whose edges are 
flushed with four of the six surface boundaries of the computational fluid domain. The free-stream velocity of the flow is perpendicular to the surface of the wall and periodic boundary conditions are 
prescribed in both transverse directions.

Assuming that the porous wall is reasonably thin, it can be modeled as a surface denoted here by $\Omega$. This surface is discretized by a mesh whose typical element size is much larger than the 
pore size. Hence, given a scalar or vector quantity $\phi$ defined in $\Omega$, a homogenized semi-discrete counterpart $\bar{\phi}$, where homogenization is performed in both transverse directions and 
designated by an overline, is defined here as
\begin{equation}
	\bar{\phi} = \displaystyle{\frac{\int_\Omega \phi \, d\Omega}{|\Omega|}}\,,
	\label{eq:averaging}
\end{equation}
where $|\Omega|$ denotes the area of the surface $\Omega$.

Due to symmetry (see \Cref{fig: computational domain}), both homogenized transverse velocity components $\bar{v}$ and $\bar{w}$ are zero. For this reason, the homogenized version of \Cref{eq: ns} 
{\it inside the porous wall} can be written as
\begin{equation}
\begin{split}
    \frac{\partial \bar\rho}{\partial t} + \frac{\partial \overline{\rho u}}{\partial x} &= 0\,,\\
    \frac{\partial \overline {\rho u}}{\partial t} + \frac{\partial \overline { \rho u  u + p}}{\partial x}  &= 
	\frac{\partial \overline {\tau_{xx}}}{\partial x} + 
	\underbrace{\left [{\frac{ \overline {\partial \tau_{xy}}}{\partial y}} + {\frac{ \overline {\partial \tau_{xz}}}{\partial z}}\right]}_{\textrm{to be modeled}},\\
    \frac{\partial \overline E}{\partial t} + \frac{\partial \overline {(E + p)u}}{\partial x} &= 
	\frac{\partial \overline{\tau_{xx} u-q_x}}{\partial x} + 
		\underbrace{\left [{\frac{\overline {\partial \tau_{xy} u}}{\partial y}} + {\frac{\overline {\partial \tau_{xz} u}}{\partial z}}\right]}_{\textrm{to be modeled}},
    \end{split}
    \label{eq: Homogenized-NS-prelim}
\end{equation}
where 
\begin{itemize}
	\item $\tau_{xx}$, $\tau_{xy}$, and $\tau_{xz}$ are the components of the viscous stress tensor ${\uptau}$ given by 
		\begin{equation} \tau_{xx} = \mu \left( \frac{4}{3}\frac{\partial u}{\partial x} - \frac{2}{3} \frac{\partial v}{\partial y}  - \frac{2}{3} \frac{\partial w}{\partial z} \right),\quad 
			\tau_{xy} = \mu \left( \frac{\partial u}{\partial y} + \frac{\partial v}{\partial x} \right),\quad \tau_{xz} = \mu \left( \frac{\partial u}{\partial z} 
			+ \frac{\partial w}{\partial x} \right). 
			\label{eq: viscous stress} 
		\end{equation} 
	\item $q_x$ is the $x$-component of the heat flux vector $\bm{q}$ whose expression is given by 
		\begin{equation} q_x = -\kappa\frac{\partial T}{\partial x}\,.
			\label{eq:qx} 
		\end{equation} 
	\item The averaged terms $\overline{\partial \tau_{xy} / \partial y} + \overline{\partial \tau_{xz} / \partial z}$ and $\overline{\partial \tau_{xy} u / \partial y} 
		+ \overline{\partial  \tau_{xz} u / \partial z}$ are shown between brackets to emphasize that: while
		they vanish in the cross sections away from the porous wall -- due to the fact that they contain only partial derivatives with respect to the directions of 
		homogenization $y$ and $z$ -- they  represent  the  averaged  friction  loads  on  the  pore boundary and therefore require subgrid scale modeling;
		and hence they will be reintroduced in \Cref{sec:BFT} in the form of a body force model.
\end{itemize}

From (\ref{eq: viscous stress}), (\ref{eq:qx}), and the periodic nature of the prescribed boundary conditions (see \Cref{fig: computational domain}), it follows that the 
expressions for the averaged, wall normal, viscous flux terms $\partial \overline {\tau_{xx}} / \partial x$ and $\partial \left( \overline{\tau_{xx} u-q_x} \right) / \partial x$ can be simplified as 
follows
\begin{equation}
	    \overline {\tau_{xx}} = \frac{4}{3}\mu\frac{\partial \overline {u}}{\partial x} \quad \textrm{ and } \quad
	     \overline{\tau_{xx} u-q_x} = \frac{4}{3}\mu\overline{\frac{\partial  u}{\partial x} u} + \kappa \frac{\partial \overline{T}}{\partial x}\,,
\end{equation}
which transforms \Cref{eq: Homogenized-NS-prelim} into the following, more specific, homogenized governing equations inside the porous wall
\begin{equation} 
	\begin{split} 
	\frac{\partial \bar\rho}{\partial t} + \frac{\partial \overline{\rho u}}{\partial x} &= 0\,,\\ 
	\frac{\partial \overline {\rho u}}{\partial t} + \frac{\partial \overline { \rho u  u + p}}{\partial x}  &= 
	\frac{\partial}{\partial x}\left(\frac{4}{3}\mu\frac{\partial \overline {u}}{\partial x}\right) + 
	\underbrace{\left [{\frac{ \overline {\partial \tau_{xy}}}{\partial y}} + {\frac{ \overline {\partial \tau_{xz}}}{\partial z}}\right]}_{\textrm{to be modeled}}\,,\\
	\frac{\partial \overline E}{\partial t} + \frac{\partial \overline {(E + p)u}}{\partial x} &= 
	\frac{\partial}{\partial x}\left(\frac{4}{3}\mu\overline{\frac{\partial  u}{\partial x} u} + \kappa \frac{\partial \overline{T}}{\partial x}\right) 
	+ \underbrace{\left [{\frac{\overline {\partial \tau_{xy} u}}{\partial y}} + {\frac{\overline {\partial \tau_{xz} u}}{\partial z}}\right]}_{\textrm{to be modeled}}.
	\end{split} 
\label{eq: Homogenized-NS}
\end{equation}

\begin{figure}[!ht]
\centering
    \centering
    \begin{tikzpicture}[scale=3]
    \draw[thick,->] (-0.9, 0.3, 0) -- (-0.4, 0.3, 0) node[anchor=north]{$x$}; 
    \draw[thick,->] (-0.9, 0.3, 0) -- (-0.9, 0.8, 0) node[anchor=south]{$y$};
    \draw[thick,->] (-0.9, 0.3, 0) -- (-0.9, 0.3, 0.5) node[anchor=south]{$z\ \ $};
    \draw [fill=white,line width=0.2mm] (0,0) rectangle (4,1);
    \draw [fill=white,line width=0.2mm] (1,0) rectangle (3,1);
    \fill [pattern = north east lines] (2 - 0.08,0) rectangle (2 + 0.08,0.4);
    \fill [pattern = north east lines] (2 - 0.08,0.6) rectangle (2 + 0.08,1.0);
     
    \node at (0.5,0.5)[circle,fill,inner sep=1.5pt]{};
    \node at (1.5,0.5)[circle,fill,inner sep=1.5pt]{};
    \node at (2.5,0.5)[circle,fill,inner sep=1.5pt]{};
    \node at (3.5,0.5)[circle,fill,inner sep=1.5pt]{};
    \draw node[scale=1.2] at (1.3,0.8) {$\overline{\bm{W}}_i$};
    \draw node[scale=1.2] at (2.7,0.8) {$\overline{\bm{W}}_j$};
    
    \draw node[scale=1.2] at (1.4,0.2) {$\mathcal C_i$};
    \draw node[scale=1.2] at (2.6,0.2) {$\mathcal C_j$};
    
    \draw node[scale=1.2] at (1.45,0.5) {$i$};
    \draw node[scale=1.2] at (2.55,0.5) {$j$};
    \draw node[scale=1.2] at (2.15,0.5) {$\partial {\mathcal C}_{ij}$ };

    \draw[thick,->,line width=0.2mm,](1.7, 0.05) -- (1.9, 0.05);
    \draw[thick,->,line width=0.2mm,](1.7, 0.15) -- (1.9, 0.15);
    \draw[thick,->,line width=0.2mm,](1.7, 0.25) -- (1.9, 0.25);
    \draw[thick,->,line width=0.2mm,](1.7, 0.35) -- (1.9, 0.35);
    \draw[thick,->,line width=0.2mm,](1.7, 0.45) -- (1.9, 0.45);
    \draw[thick,->,line width=0.2mm,](1.7, 0.55) -- (1.9, 0.55);
    \draw[thick,->,line width=0.2mm,](1.7, 0.65) -- (1.9, 0.65);
    \draw[thick,->,line width=0.2mm,](1.7, 0.75) -- (1.9, 0.75);
    \draw[thick,->,line width=0.2mm,](1.7, 0.85) -- (1.9, 0.85);
    \draw[thick,->,line width=0.2mm,](1.7, 0.95) -- (1.9, 0.95);
    
    \draw [red, thick, dashed] (1,0)   rectangle (2, 0.4);
    \draw [red, thick, dashed] (1,0.6) rectangle (2, 1.0);
    \draw [cyan, thick, dashed] (1,0.4) rectangle (2, 0.6);
    
    \draw [red, thick, dashed] (2,0)   rectangle (3, 0.4);
    \draw [red, thick, dashed] (2,0.6) rectangle (3, 1.0);
    \draw [cyan, thick, dashed] (2,0.4) rectangle (3, 0.6);
    
    \end{tikzpicture}
	\caption{Schematic of the homogenized inviscid flux: context of a cell-centered finite volume approximation method on a body-fitted CFD mesh (central $x$-$y$ plane view). }
\label{fig: homogenized flux}
\end{figure}
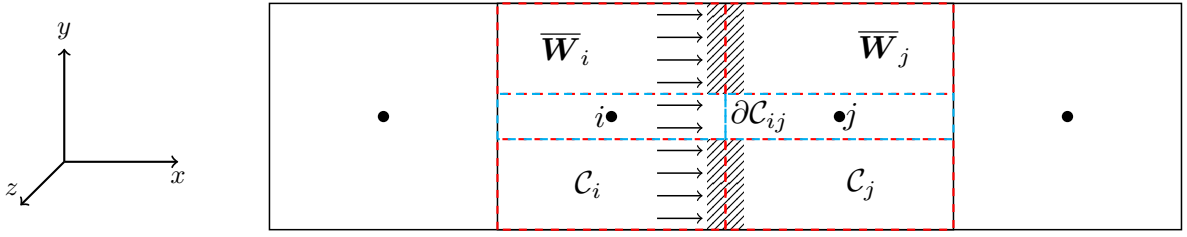

For the sake of clarity, the homogenization approach proposed herein is described in the next two subsections not only in the backdrop of the model porous flow problem illustrated in 
\Cref{fig: computational domain} but also in the context of a cell-centered Finite Volume (FV) approximation method, where the gradients underlying the viscous fluxes are 
semi-discretized by a finite difference scheme and for simplicity, but without any loss of generality, the CFD mesh is assumed to be structured and body-fitted. In this case, each computational cell 
${\mathcal C}_i$ is a primal element of the CFD mesh, $i$ is the center point of ${\mathcal C}_i$ where the semi-discrete fluid state vector is defined, and the boundary surface (cell facet) 
$\partial {\mathcal C}_{ij}$ shared by two cells $\mathcal C_i$ and $\mathcal C_j$, where $j$ is such that the line $(i, j)$ traverses the porous wall, lies along the surface of this wall 
(see \Cref{fig: homogenized flux}). However, as will be noted wherever appropriate within the next two subsections, the generalizations of this approach to a vertex-based FV approximation method 
where the viscous fluxes are semi-discretized by a Finite Element (FE) method and to unstructured meshes as well as an arbitrary orientation of the porous wall with respect to the global reference 
frame, are straightforward. Furthermore, the homogenization approach described below is not limited to body-fitted CFD methods: it is equally applicable to embedded/immersed boundary methods. In 
particular, it is well-suited for the FInite Volume method with Exact two-material Riemann problems (FIVER) \cite{wang2011algorithms, lakshminarayan2014embedded,main2017enhanced, huang2018family, 
borker2019mesh} (see also \Cref{sec: implementation}). As a matter of fact, all numerical results with the porosity model reported in \Cref{sec: app} are obtained using non body-fitted CFD meshes and the Embedded Boundary 
Method (EBM) FIVER. In short, the proposed homogenization approach for the treatment of porous wall boundary conditions can be incorporated in any semi-discretization or discretization method.

\vglue 0.30truein

\subsection{Homogenized inviscid flux}\label{sec: inviscid}

Let $\alpha$ denote the void fraction or porosity of the porous wall. It can be expressed as
\begin{equation}
	\alpha = \displaystyle{\frac{|\Omega|^{\textrm{pore}}}{|\Omega|}}
	\label{eq:alpha}
\end{equation}
where $|\Omega^{\textrm{pore}}|$ denotes the total area occupied by the pores in the surface $\Omega$.

Consider first the two following extreme cases: 
\begin{itemize}
	\item $\alpha = 1$, which corresponds to the no-wall configuration. In this case, the inviscid fluxes through the (nonexistent) porous wall can be approximated as usual, for example, by
		an approximate Riemann solver such as Roe's solver \cite{roe1981approximate}. This can be written as (see \Cref{fig: homogenized flux})
		\begin{equation}
			\bm{F}_{ij} = - \bm{F}_{ji} = |{\partial \mathcal C}_{ij}|\,\bm{F}^{\textrm{\textrm{Roe}}}(\bm{W}_i, \bm{W}_j, n_{ij})\,,
			\label{eq:ftpw}
		\end{equation}
		where $\bm{F}_{ij}$ is the numerical inviscid flux function associated with the inviscid flux vector in the $x$-direction, $\bm{F}$, $n_{ij}$ is the unit outward normal to 
		$\partial {\mathcal C}_{ij}$ -- in this case, $n_{ij}$ is parallel to the $x$-direction -- and $|\partial{\mathcal C}_{ij}|$ denotes the area of $\partial{\mathcal C}_{ij}$.
	\item $\alpha = 0$, which corresponds to the impermeable wall configuration. In this case, the inviscid fluxes can also be approximated as usual -- that is, as wall boundary fluxes as follows 
		(see \Cref{fig: homogenized flux})
		\begin{equation} 
			\bm{F}_{ij} = |\partial{\mathcal C}_{ij}|\,\bm{F}^{\textrm{\textrm{imp-wall}}}(\bm{W}_i, n_{ij}) \quad \textrm{ and } \quad \bm{F}_{ji} = |\partial{\mathcal C}_{ji}|\,\bm{F}^{\textrm{\textrm{imp-wall}}}(\bm{W}_j, n_{ji})\,,
			\label{eq:ftiw}
		\end{equation}
\end{itemize}
where $|\partial{\mathcal C}_{ji}| = |\partial{\mathcal C}_{ij}|$, $n_{ji} = - n_{ij}$, and $\bm{F}^{\textrm{\textrm{imp-wall}}}(\bm{W}_i, n_{ij})$ and 
$\bm{F}^{\textrm{\textrm{imp-wall}}}(\bm{W}_j, n_{ji})$ are computed using the physical inviscid flux tensor $\mathcal F$, $n_{ij}$, and the standard, {\it impermeable wall} boundary conditions.

Hence, the main idea here is to compute, for any nontrivial porosity value $ 0 < \alpha < 1$, the homogenized, numerical inviscid flux functions through the porous wall $\bm{F}^{\textrm{ave}}_{ij}$ and 
$\bm{F}^{\textrm{ave}}_{ji}$ as convex combinations of \eqref{eq:ftpw} and \eqref{eq:ftiw} using however averaged values of the fluid state vectors -- that is,
\begin{equation}
\begin{split}
	\bm{F}^{\textrm{ave}}_{ij} &= |\partial{\mathcal C}_{ij}|\left(\alpha \bm{F}^{\textrm{\textrm{Roe}}}(\overline{\bm{W}}_i,\overline{\bm{W}}_j, n_{ij}) + (1 - \alpha) \bm{F}^{\textrm{imp-wall}}(\overline{\bm{W}}_i, n_{ij})\right)\,,\\
	\bm{F}^{\textrm{ave}}_{ji} &= |\partial{\mathcal C}_{ji}|\left(\alpha \bm{F}^{\textrm{\textrm{Roe}}}(\overline{\bm{W}}_j,\overline{\bm{W}}_i, n_{ji}) + (1 - \alpha) \bm{F}^{\textrm{imp-wall}}(\overline{\bm{W}}_j, n_{ji})\right)\,,
\end{split}
\label{eq: homogeneous inviscid flux}
\end{equation}
where averaging is performed as in \eqref{eq:averaging}. The second-order extension of the above approximations using the classical Monotonic Upstream-centered Scheme for Conservation Laws 
(MUSCL) is straightforward.

The above computation \eqref{eq: homogeneous inviscid flux} is illustrated in \Cref{fig: homogenized flux}, where the computational cells adjacent to the porous wall are divided into two parts: one part 
attached to a hole (colored in blue) representing the pores, collectively; and another part attached to the wall (colored in red). Accordingly, the flux through the porous wall $\bm{F}^{\textrm{ave}}_{ij}$
consists of the flux through the hole \eqref{eq:ftpw} and that at the impermeable part of the wall (see \Cref{eq:ftiw}). 

Note that by construction, the proposed homogenization of the numerical inviscid flux function \eqref{eq: homogeneous inviscid flux} associated with the inviscid flux vector in the $x$-direction $\bm{F}$
conserves mass and energy; it also guarantees that the impermeable part of the numerical flux has zero mass and energy components.

The generalization of the inviscid flux homogenization procedure described above to the case of a vertex-based FV method with control volumes (dual cells) ${\mathcal C}_i$ -- in which case 
a wall boundary grid point $i$ and not a wall boundary cell facet $\partial {\mathcal C}_{ij}$ lies on the porous wall -- is simply accomplished using the concept of half dual cells. The generalization 
of this procedure to an arbitrary direction of the outward normal to the wall $n_{ij}$ and to unstructured meshes is straightforward, because most if not all approximate Riemann solvers are essentially 
one-dimensional local solvers that propagate information in the direction normal to the considered facet of the control volume boundary. 

\subsection{Homogenized viscous flux}\label{sec: viscous}

The spirit of the approach described in \Cref{sec: inviscid} for the numerical inviscid fluxes is followed here to model each numerical viscous flux in 
the wall normal direction (which in this case is the $x$-direction), $\bm{G}_{ij}^{\textrm{ave}}$ and $\bm{G}_{ji}^{\textrm{ave}}$, as a convex combination of the numerical viscous flux 
at an impermeable surface and that through pores computed using averaged fluid state vectors. Recalling that the terms shown between brackets in \Cref{eq: Homogenized-NS} vanish
because they contain derivatives in the homogenized directions, choosing to approximate the gradients in the viscous fluxes by, for example, the second-order central finite differencing, and assuming
that the dynamic shear viscosity $\mu$ and fluid thermal conductivity $\kappa$ are constant, the homogenization of the numerical viscous flux functions through the porous wall can be carried out as 
follows (see \Cref{fig: homogenized flux})

\begin{equation}
\begin{split}
	\bm{G}_{ij}^{\textrm{ave}} = |\partial{\mathcal C}_{ij}|\left(\alpha \underbrace{\begin{pmatrix}
           0 \\
           \frac{4}{3}\mu\frac{\overline {u}_j - \overline {u}_i}{\delta_x}\\
           \frac{4}{3}\mu \left( \frac{\overline {u}_j - \overline {u}_i}{\delta_x} \right) \left( \frac{\overline {u}_i + \overline {u}_j}{2} \right) + \kappa \frac{\overline{T}_j - \overline{T}_i}{\delta_x}  
	\end{pmatrix}}_{\bm{G}_{ij}^{\textrm{pore}}}
	 \,+\, 
	 (1 - \alpha) \underbrace{\begin{pmatrix}
           0 \\
           \frac{4}{3}\mu\frac{\overline {u}^{\,\textrm{be}}_j - \overline {u}_i}{\delta_x} \\
		 \frac{4}{3}\mu \left( \frac{\overline {u}^{\,\textrm{be}}_j - \overline {u}_i}{\delta_x} \right) \left( \frac{\overline {u}_i + \overline {u}^{\,\textrm{be}}_j}{2} \right) + \kappa \frac{\overline{T}^{\,\textrm{be}}_j - \overline{T}_i}{\delta_x}  
	 \end{pmatrix}}_{\bm{G}_{ij}^{\textrm{imp-wall}}}\right)\,,\\
	\bm{G}_{ji}^{\textrm{ave}} = |{\partial\mathcal C}_{ij}|\left(\alpha \underbrace{\begin{pmatrix}
           0 \\
           \frac{4}{3}\mu\frac{\overline {u}_i - \overline {u}_j}{\delta_x} \\
           \frac{4}{3}\mu \left( \frac{\overline {u}_i - \overline {u}_j}{\delta_x} \right) \left( \frac{\overline {u}_i + \overline {u}_j}{2} \right) + \kappa \frac{\overline{T}_i - \overline{T}_j}{\delta_x}  
	\end{pmatrix}}_{\bm{G}_{ji}^{\textrm{pore}}}
         \,+\,
	 (1 - \alpha) \underbrace{\begin{pmatrix}
           0 \\
           \frac{4}{3}\mu\frac{\overline {u}^{\,\textrm{be}}_i - \overline {u}_j}{\delta_x} \\
		 \frac{4}{3}\mu \left( \frac{\overline {u}^{\,\textrm{be}}_i - \overline {u}_j}{\delta_x} \right) \left( \frac{\overline {u}^{\,\textrm{be}}_i + \overline {u}_j}{2} \right) + \kappa \frac{\overline{T}^{\,\textrm{be}}_i - \overline{T}_j}{\delta_x}  
	 \end{pmatrix}}_{\bm{G}_{ji}^{\textrm{imp-wall}}}\right)\,,
\end{split}
\label{eq: homogeneous viscous flux}
\end{equation}
where $\delta_x$ denotes the mesh spacing in the $x$-direction and $\overline{u}^{\,\textrm{be}}_i$, $\overline{u}^{\,\textrm{be}}_j$, $\overline{T}^{\,\textrm{be}}_i$, 
and $\overline{T}^{\,\textrm{be}}_j$ are nodal velocity and temperature values populated using any preferred extrapolation approaches that enforce the no-slip and adiabatic wall boundary conditions in 
cell-centered FV approximation methods.

As stated earlier, the vanishing terms $\overline{\partial \tau_{xy} / \partial y} + \overline{\partial \tau_{xz} / \partial z}$ and $\overline{\partial \tau_{xy} u / \partial y} + \overline{\partial  
\tau_{xz} u / \partial z}$ of \Cref{eq: Homogenized-NS} must be modeled as they represent the averaged friction loads (per unit volume) at a pore boundary (or at an expanded jet stream boundary). 
The evaluation of these terms requires the availability of a profile of the component of the flow velocity vector normal to the porous wall -- in this case, $u(y, z)$ -- that is fully resolved inside a 
pore: hence, it requires subgrid scale modeling.
Such modeling is inspired here by the Hagen-Poiseuille flow \cite{sutera1993history}, which in this case is homogeneous in the $x$-direction. It starts by postulating: an equivalent
pore -- that is, a single pore whose shape and dimensions are such that when inserted at the center of a unit cell, the void fraction of the cell matches that of the fabric of interest; and a parabolic 
shape for the profile of the wall normal velocity component in the equivalent pore that depends on its geometry. The subgrid scale modeling ends by deriving an analytical expression 
of the form of a body force term for the homogenized equations of dynamic equilibrium (\ref{eq: Homogenized-NS}). Then, the body force term, which represents the friction loss along pore boundaries, is 
approximated by any preferred numerical procedure.

\begin{figure}[!ht]
\centering
\begin{subfigure}[b]{0.06\textwidth}
    \centering
    \begin{tikzpicture}[scale=3]
      \draw[thick,->] (0.5,0.7) -- (0.8,0.7) node[anchor=north]{$y$}; 
      \draw[thick,->] (0.5,0.7) -- (0.5,1.0) node[anchor=south]{$z$};
    \end{tikzpicture}
    \caption*{ }
\end{subfigure}
\begin{subfigure}[b]{0.2\textwidth}
    \centering
    \begin{tikzpicture}[scale=3]
      \fill[cyan] (0,0) rectangle (1.0,1.0);
      \fill[white] (0.5,0.5) circle (0.2);
      \draw[thick,->] (0.5,0.5) -- (0.7,0.5) node[midway , above=0.1]{$r$};
    \end{tikzpicture}
    \caption{circular} \label{fig: geometry0_circle}
\end{subfigure}
\begin{subfigure}[b]{0.2\textwidth}
  \centering
   \begin{tikzpicture}[scale=3]
      \fill[cyan] (0,0) rectangle (1.0,1.0);
      \fill[white] (0.3,0.3) rectangle (0.7,0.7);
      \draw[thick,->] (0.5,0.5) -- (0.7,0.5) node[midway , above=0.1]{$r$};
    \end{tikzpicture}
    \caption{square} \label{fig: geometry0_square}
\end{subfigure}
\begin{subfigure}[b]{0.2\textwidth}
  \centering
   \begin{tikzpicture}[scale=3]
      \fill[cyan] (0,0) rectangle (0.4,1.0);
      \fill[cyan] (0.6,0) rectangle (1.0,1.0);
      \draw[thick,->] (0.5,0.5) -- (0.6,0.5) node[midway , above=0.1]{$r$};
    \end{tikzpicture}
    \caption{gapped} \label{fig: geometry0_gap}
\end{subfigure}
\caption{Idealized symmetric pore geometries.}
\label{fig: geometry0}
\end{figure}
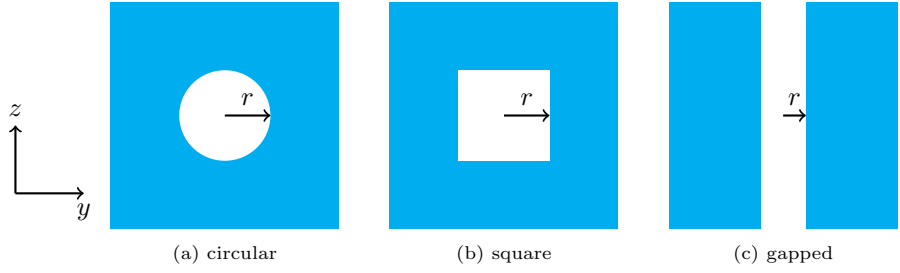

To this end, circular, square, and gapped shapes pores (see \Cref{fig: geometry0}) are examined next. For each of these geometries, which can be considered as an idealization of a family of more complex 
geometries, a velocity profile $u(y, z)$ inside the pore is postulated and the form of the corresponding body force term is derived. 

At this point, it is noted that the extension of the homogenization of the numerical viscous flux functions (\ref{eq: homogeneous viscous flux}) to the case of a vertex-based FV 
approximation method, where the viscous fluxes are semi-discretized by a FE method, is straightforward. In such a context, the computational cell ${\mathcal C}_i$ becomes the dual cell associated with 
the grid point $i$ of the CFD mesh, which can be structured or arbitrarily unstructured. If this mesh is body-fitted, all wall boundary grid points lie on the wall boundary. In this subcase,
$|{\mathcal C}_i|$ is replaced by the volume of a primal element $B^e$ of the CFD mesh that is connected to the grid point $i$; the computation of the term ${\bm{G}_{ij}^{\textrm{pore}}}$ is unchanged 
except that the constant gradients in this term are computed by differentiating the linear shape functions of the element $B^e$; and summation is performed over all elements $B^e$ connected to the grid 
point $i$. As for the term ${\bm{G}_{ij}^{\textrm{imp-wall}}}$, it simplifies to the standard computation of the wall boundary viscous fluxes -- that is, no extrapolation is needed; and similarly, 
summation is performed over all elements $B^e$ connected to the grid point $i$. On the other hand, if the CFD mesh is not body-fitted, a preferred embedded/immersed boundary method can be used as is or 
tailored to compute both terms ${\bm{G}_{ij}^{\textrm{pore}}}$ and ${\bm{G}_{ij}^{\textrm{imp-wall}}}$ (for example, see \Cref{sec: implementation}), whether the FV approximation method is cell-centered 
or vertex-based.

\subsubsection{Circular equivalent pore}
\label{sec:CEP}

For a circular equivalent pore such as that shown in \Cref{fig: geometry0_circle}, the Hagen-Poiseuille flow theory leads to the following velocity profile inside the pore 
\cite{white2006viscous}
\begin{equation}\label{eq: cicle velocity}
    u(y, z) = C \left( r^2 - y^2 - z^2 \right),
\end{equation}
where $r$ is the radius of the circular geometry and $C$ is a free parameter that will be shown not play an explicit role in the computation of the body force term. 
From (\ref{eq:averaging}), (\ref{eq:alpha}), and (\ref{eq: cicle velocity}),
it follows that in this case, the averaged flow velocity is
\begin{equation}
	\overline{u} = \displaystyle{\frac{\int_\Omega u \, dy dz}{|\Omega|} = \frac{\int_{\Omega^{\textrm{pore}}} u \, dy dz}{|\Omega|}  = C \frac{|\Omega^{\textrm{pore}}|}{|\Omega|}\frac{r^2}{2} 
	= C\alpha \frac{r^2}{2}}\,.
    \label{eq: cicle velocity ave}
\end{equation}

From (\ref{eq: cicle velocity}) and (\ref{eq: cicle velocity ave}), it follows that
\begin{equation}
\begin{split}
    \overline{\frac{\partial^2 u}{\partial y^2} + \frac{\partial^2 u}{\partial z^2}} &= -8 \frac{\overline{u}}{r^2}\,,\\
    \overline{u \left( \frac{\partial^2 u}{\partial y^2} + \frac{\partial^2 u}{\partial z^2} \right) } &= -8 \frac{\overline{u}^2}{\alpha r^2}\,,\\
    \overline{ \left( \frac{\partial u}{\partial y} \right)^2 + \left( \frac{\partial u}{\partial z} \right)^2} &= 8 \frac{\overline{u}^2}{\alpha r^2}\,.
\end{split}
\label{eq: circle visc}
\end{equation}

Substituting the results (\ref{eq: circle visc}) into the expressions (\ref{eq: viscous stress}) leads to the following subgrid scale model for the friction loss along circular pore boundaries

\begin{equation}
\begin{split}
	\displaystyle{\overline{\frac{\partial \tau_{xy}}{\partial y}} + \overline{\frac{\partial \tau_{xz}}{\partial z}}} &= -8\mu \frac{\overline{u}}{r^2}\,,\\
	\displaystyle{\overline{\frac{\partial \tau_{xy} u}{\partial y}} + \overline{\frac{\partial  \tau_{xz} u}{\partial z}}} &= 0\,.
    \end{split}\label{eq: circle dissipation}
\end{equation}

\subsubsection{Square equivalent pore}
\label{sec:SEP}

For a square equivalent pore such as that shown in \Cref{fig: geometry0_square}, the Hagen-Poiseuille flow theory leads to the following velocity profile inside the pore 
\cite[p. 113]{white2006viscous}

\begin{equation}
	u(y, z) = \displaystyle{C\frac{16r^2}{ \pi^3}\sum_{i=1,3,5 ...}^{\infty} \left[ (-1)^{(i-1)/2} \left( 1 - \frac{\cosh{(i\pi z/2r)}}{\cosh{(i\pi/2)}} \right) \frac{\cos{(i\pi y/2r)}}{i^3} \right],}
    \label{eq: square velocity}
\end{equation}
where $2r$ is the side length of the square geometry and $C$ is a free parameter that will be shown not play an explicit role in the computation of the body force term. 
From (\ref{eq:averaging}), (\ref{eq:alpha}), and (\ref{eq: square velocity}), it follows that in this case, 
the averaged flow velocity is given by
\begin{eqnarray}\label{eq: square velocity ave}
	\overline{u} &=& \displaystyle{\frac{\int_\Omega u \, dy dz}{|\Omega|} = \frac{\int_{\Omega^{\textrm{pore}}} u \, dy dz}{|\Omega|} = 
	C \frac{|\Omega^{\textrm{pore}}|}{|\Omega|}\frac{r^{2}}{3} \left(1 - \frac{192}{\pi^5}\sum_{i=1,3,5 ...}^{\infty}\frac{\tanh{(i\pi/2)}}{i^5} \right)} \nonumber\\
	&=& C \alpha \displaystyle{\frac{r^{2}}{3} \left(1 - \frac{192}{\pi^5}\sum_{i=1,3,5 ...}^{\infty}\frac{\tanh{(i\pi/2)}}{i^5} \right)}\,.
\end{eqnarray}

From (\ref{eq: square velocity}) and (\ref{eq: square velocity ave}), it follows that
\begin{equation}
\begin{split}
    \overline{\frac{\partial^2 u}{\partial y^2} + \frac{\partial^2 u}{\partial z^2}} &= - \frac{3}{\zeta} \frac{\overline{u}}{r^2}\,,\\
    \overline{u \left( \frac{\partial^2 u}{\partial y^2} + \frac{\partial^2 u}{\partial z^2} \right) }  + 
    \overline{\left( \frac{\partial u}{\partial y}\Big)^2 + \Big(\frac{\partial u}{\partial z} \right)^2} &= 0\,,
\end{split}\label{eq: square visc}
\end{equation}
where $$\zeta =  1 - \frac{192}{\pi^5}\sum_{i=1,3,5 ...}^{\infty}\frac{\tanh{(i\pi/2)}}{i^5} \approx 0.421731044865.$$

Substituting (\ref{eq: square visc}) into (\ref{eq: viscous stress}) leads to the following subgrid scale model for the friction loss along the boundaries of a square pore
\begin{equation}
\begin{split}
    \overline{\frac{\partial \tau_{xy}}{\partial y}} + \overline{\frac{\partial \tau_{xz}}{\partial z}} &= -\frac{3}{\zeta} \mu \frac{\overline{u}}{r^2}\,,\\
     \overline{\frac{\partial \tau_{xy} u}{\partial y}} + \overline{\frac{\partial  \tau_{xz} u}{\partial z}} &= 0\,.
    \end{split}\label{eq: square dissipation}
\end{equation}

\subsubsection{Gapped shape equivalent pore}
\label{sec:GSEP}

For the gapped shape equivalent pore shown in \Cref{fig: geometry0_gap}, the Hagen-Poiseuille flow theory leads to the following velocity profile inside the pore \cite{white2006viscous}
\begin{equation}\label{eq: slot velocity}
    u(y, z) = C \left( r^2 - y^2 \right),
\end{equation}
where $2r$ is in this case the width of the gap and $C$ is a free parameter that will be shown not to play an explicit role in the computation of the body force term. 
From (\ref{eq:averaging}), (\ref{eq:alpha}), and (\ref{eq: slot velocity}, it follows that in this case,
the averaged flow velocity can be written as
\begin{equation}\label{eq: slot velocity ave}
	\overline{u} = \displaystyle{\frac{\int_\Omega u \, dy dz}{|\Omega|} = \frac{\int_{\Omega^{\textrm{pore}}} u \, dy dz}{|\Omega|} = C \frac{|\Omega^{\textrm{pore}}|}{|\Omega|}\frac{2r^{2}}{3} =
	C \alpha \frac{2r^2}{3}}.
\end{equation}

From (\ref{eq: slot velocity}) and (\ref{eq: slot velocity ave}), it follows that
\begin{equation}
\begin{split}
    \overline{\frac{\partial^2 u}{\partial y^2} + \frac{\partial^2 u}{\partial z^2}} &= -3 \frac{\overline{u}}{ r^2}\,,\\
    \overline{u \left(\frac{\partial^2 u}{\partial y^2} + \frac{\partial^2 u}{\partial z^2} \right) } &= -3 \frac{\overline{u}^2}{\alpha r^2}\,,\\
    \overline{\left(\frac{\partial u}{\partial y} \right)^2 + \left(\frac{\partial u}{\partial z} \right)^2} &= 3  \frac{\overline{u}^2}{\alpha r^2}\,.
\end{split}\label{eq: slot visc}
\end{equation}

Substituting the results (\ref{eq: slot visc}) into (\ref{eq: viscous stress}) yields the following subgrid scale model for the friction loss along the boundaries of a gapped shape pore
\begin{equation}
\begin{split}
    \overline{\frac{\partial \tau_{xy}}{\partial y}} + \overline{\frac{\partial \tau_{xz}}{\partial z}} &= -3 \mu \frac{\overline{u}}{r^2}\,,\\
     \overline{\frac{\partial \tau_{xy} u}{\partial y}} + \overline{\frac{\partial  \tau_{xz} u}{\partial z}} &= 0\,.
    \end{split}\label{eq: slot dissipation}
\end{equation}

\subsection{Body force term}
\label{sec:BFT}

From the second of each of Eqs. (\ref{eq: circle dissipation}), (\ref{eq: square dissipation}), and (\ref{eq: slot dissipation}), it follows that whatever symmetric pore geometry is assumed, 
the resulting friction loss contributes only to a momentum loss. In particular, it does not contribute to a total energy loss, because friction converts the lost kinetic energy into internal energy. 
For $\alpha = 8\%$, the momentum loss for a circular pore with $\displaystyle r = \sqrt{{\alpha |\Omega|}/{\pi}}$ represented by the form of the body force term derived in 
(\ref{eq: circle dissipation})  is about $\displaystyle -314.16 \mu (\overline{u}/|\Omega|)$, that for a square pore with $\displaystyle r = \sqrt{{\alpha |\Omega|}/{4}}$ represented by the counterpart 
form (\ref{eq: square dissipation}) is roughly $\displaystyle -355.68 \mu ( \overline{u}/|\Omega|)$, and that due to a gapped shape pore with $\displaystyle r = {(\alpha \sqrt{|\Omega|})}/{2}$ and the 
form of its body force term (\ref{eq: slot dissipation}) is around $\displaystyle -1875 \mu (\overline{u}/|\Omega|)$. This observation suggests that the smaller is the aspect ratio of a pore, the 
smaller is the associated momentum loss or dissipation.

In any case, the derived form of the body force term is multiplied by the following Gaussian approximation of the Dirac function in order to locate it along the pore boundaries
\begin{equation}
    D(x) = \exp{\left[ -\frac{1}{2} \left(\frac{x - x_0}{\sigma} \right)^2 \right] }\,,
    \label{eq: source dist}
\end{equation}
where $x_0$ is the position of the porous wall -- in this case along the $x$-direction -- 
$\sigma = \eta_f/(2\pi)$, and $\eta$ is the thickness of the porous wall (note that the integral of $D(x)$ along the thickness of a porous wall
is equal to the thickness of the wall). Then, the following body force (source) term is added to the homogenized equations of dynamic equilibrium (\ref{eq: Homogenized-NS})
\begin{equation}
	\bm{S} (x) = \begin{pmatrix} 0 \\
      \eta_f D(x) \left( \overline{\frac{\partial \tau_{xy}}{\partial y}} + \overline{\frac{\partial \tau_{xz}}{\partial z}} \right) \\
      0 \end{pmatrix}\,,
      \label{eq: source term}
\end{equation}
where
\begin{itemize}
	\item $\eta_f$ is a thickness correction factor that transforms the thickness of the porous wall into an ``effective'' thickness in order to account for the expanded jet stream through a pore, 
		which also contributes to the momentum loss.
	\item The term $( \overline{\partial \tau_{xy}/\partial y} + \overline{\partial \tau_{xz}/\partial z} )$ is given in the first of Eqs. (\ref{eq: circle dissipation})
		in the case of a circular pore, the first of Eqs. (\ref{eq: square dissipation}) in the case of a square pore, and the first of Eqs. (\ref{eq: slot dissipation}) in the case of
		a gapped-shape pore.
	\item In each of the first of Eqs. (\ref{eq: circle dissipation}), the first of Eqs. (\ref{eq: square dissipation}), and the first of Eqs. (\ref{eq: slot dissipation}), $\bar u$ is {\it not}
		computed using (\ref{eq: cicle velocity ave}), (\ref{eq: square velocity ave}), and (\ref{eq: slot velocity ave}), respectively. Instead, $\bar u$ is explicited computed using the
		averaging definition~(\ref{eq:averaging}). Consequently, the free parameter
		$C$ in (\ref{eq: cicle velocity}), (\ref{eq: square velocity}), and (\ref{eq: slot velocity}) is never used explicitly in the computation of the body force term (\ref{eq: source term}).
\end{itemize}

\section{Demonstration and Verification}\label{sec: app}

\begin{figure}[!ht]
  \centering
  \includegraphics[width=0.4\linewidth]{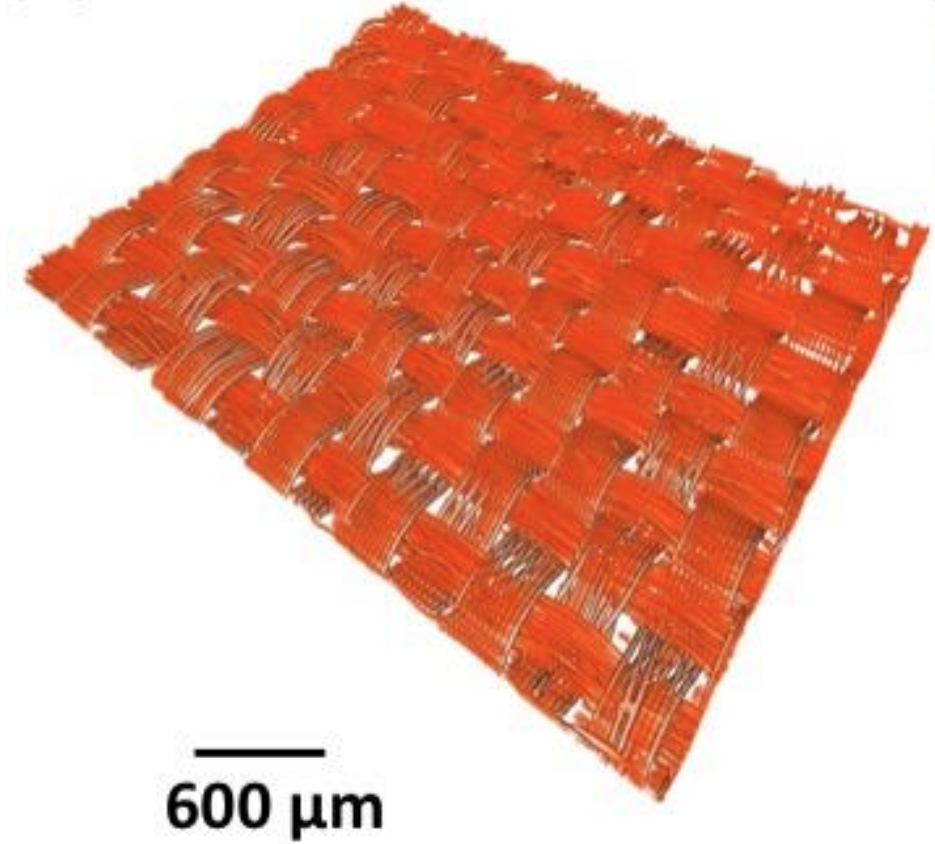}
	\caption{Microtomography of a parachute fabric (image credit: NASA/LBNL-Berkeley \cite{paneraix}).}
  \label{fig: fabric}
\end{figure}

As stated above, the main interest here is in compressible and particularly supersonic flows through permeable surfaces. A notable application is the prediction of parachute inflation dynamics
during Mars landing operations~\cite{huang2018simulation, borker2019mesh, huang2020embedded}. For this application, a relevant fabric is the PIA-C-7020 Type I fabric with Ripstop weave selected
for the parachute canopy of the Mars Science Laboratory. Based on the X-ray microtomography results published in \cite{paneraix}, the porosity of this fabric (see \Cref{fig: fabric}) is estimated
at roughly 8\%. Its thickness is around $80\ \mathrm{\mu m}$ \cite{lin2010flexible}. In both in-plane directions, the period of the weave pattern is about $500\ \mathrm{\mu m}$.

To the best of knowledge of the authors, only low-speed permeability test data is available for the PIA-C-7020 Type I fabric with Ripstop weave (and for that matter, any other fabric material). Hence, 
the proposed homogenization approach for the treatment of porous wall boundary conditions is demonstrated and verified here for supersonic flows as follows:
\begin{itemize}
	\item First, a most-suitable equivalent pore for the complex Ripstop weave pattern is determined by exploring the potential of various candidate pore shapes. For each such candidate: a series of
		pore-resolved, low-speed, unsteady numerical simulations is performed; and the computed variation of the averaged velocity through the fabric with the difference between the prescribed
		upstream and downstream pressure values, which defines the permeability of the fabric, is compared to its counterpart obtained from available low-speed permeability test data. Among all
		explored candidate pore geometries, the most-suitable one is identified as that which leads to the closest value of the experimentally measured permeability of the fabric.
	\item Next, supersonic, pore-resolved, unsteady numerical simulations are performed for the identified most-appropriate pore geometry and for another test geometry in order to generate 
		high-speed reference flow results in lieu of high-speed permeability test data. Counterpart unsteady simulations using the proposed homogenization approach are also carried out. 
		The obtained numerical results are compared to their reference values.
\end{itemize}

\subsection{General problem setup}
\label{sec:SETUP}

Given the periodicity of the weave pattern, all aforementioned numerical simulations are performed at the coupon level. Specifically, all unsteady simulations are carried out for a small square piece of 
fabric of edge size equal to $500\ \mathrm{\mu m}$, which corresponds to the period of the weave pattern. For simplicity, all flexibility effects of the coupon are neglected. For the pore-resolved
numerical simulations, the weave pattern is simplified by lumping all gaps between spun fibers into a single square or rectangular hole that is placed at the center of the coupon and sized so that the 
void fraction $\alpha$ of the material is preserved. 

\begin{figure}[!ht]
\centering
\begin{subfigure}[b]{0.06\textwidth}
    \centering
    \begin{tikzpicture}[scale=3]
      \draw[thick,->] (0.5,0.7) -- (0.8,0.7) node[anchor=north]{$y$}; 
      \draw[thick,->] (0.5,0.7) -- (0.5,1.0) node[anchor=south]{$z$};
    \end{tikzpicture}
    \caption*{ }
\end{subfigure}
\begin{subfigure}[b]{0.2\textwidth}
    \centering
    \begin{tikzpicture}[scale=3]
      \fill[cyan] (0,0) rectangle (1.0,1.0);
      \fill[white] (2.5/7.,2.5/7.) rectangle (4.5/7.,4.5/7.0);
    \end{tikzpicture}
    \caption{$2/7 \times 2/7$}
\end{subfigure}
\begin{subfigure}[b]{0.2\textwidth}
  \centering
   \begin{tikzpicture}[scale=3]
      \fill[cyan] (0,0) rectangle (1.0,1.0);
      \fill[white] (0.4,0.3) rectangle (0.6,0.7);
    \end{tikzpicture}
    \caption{$20\% \times 40\%$}
\end{subfigure}
\begin{subfigure}[b]{0.2\textwidth}
    \centering
    \begin{tikzpicture}[scale=3]
      \fill[cyan] (0,0) rectangle (1.0,1.0);
      \fill[white] (0.45, 0.1) rectangle (0.55, 0.9);
    \end{tikzpicture}
    \caption{$10\% \times 80\%$}
\end{subfigure}
\begin{subfigure}[b]{0.2\textwidth}
  \centering
   \begin{tikzpicture}[scale=3]
      \fill[cyan] (0,0) rectangle (0.46,1.0);
      \fill[cyan] (0.54,0) rectangle (1.0,1.0);
    \end{tikzpicture}
    \caption{$8\% \times 100\%$}
\end{subfigure}
	\caption{Simple equivalent pore candidates with $\alpha = 8\%$ but different aspect ratios: (a) 1:1; (b) 2:1; (c) 8:1; and (d) 12.5:1.}
\label{fig: geometry}
\end{figure}
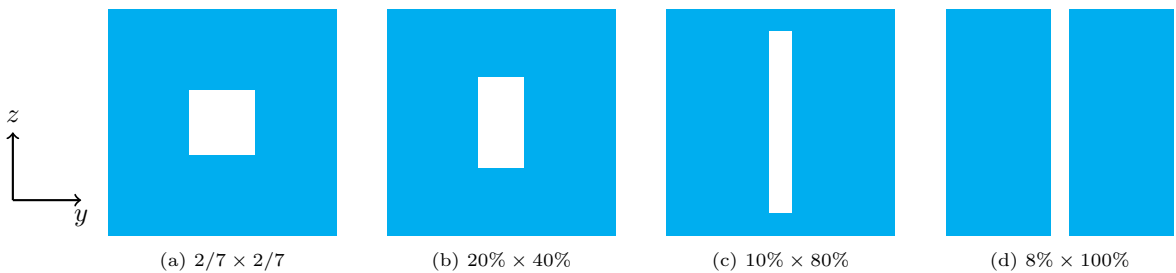

The explored equivalent pores are shown in \Cref{fig: geometry}. All have the same porosity $\alpha = 8\%$, but different aspect ratios -- specifically, $2/7\times2/7$, 
$20\%\times40\%$, $10\%\times 80\%$, and $8\% \times 100\%$. The fabric coupon is placed at the center of the computational fluid domain, as in \Cref{fig: computational domain}.
Periodic boundary conditions are prescribed in both transverse directions, but various initial and uniform upstream/downstream boundary conditions are considered in order to mimic the experimental setup 
described next.

\newpage
\subsection{Exploration of candidate equivalent pores}\label{sec: exp}

The permeability of the PIA-C-7020 Type I fabric with Ripstop weave was measured and reported in \cite{cruz2017permeability}. A series of tests were conducted using a Textest Instruments FX 3300 
Labotester III Air Permeability Tester, which consists of a test chamber with a vacuum pump. In all tests, the fabric sample was clamped over the test head opening and the upstream flow pressure was set 
to $p = 94,625 \ \mathrm{Pa}$. The temperature of the laboratory where the experiment was conducted was measured to be $T = 296.5 \ \mathrm{K}$ and the following air properties were assumed
\begin{equation}
   \gamma = 1.4, \quad R = 287.2 \ \mathrm{J \ kg^{-1}K^{-1}},\quad \textrm{and} \quad \mu = 1.846\times 10^{-5} \  \mathrm{kg \ m^{-1}s^{-1}}.
\end{equation}
Each different test was performed using still air and a different downstream pressure value that was maintained constant in time using the vacuum pump in the chamber. Then, the permeability of the fabric 
material was deduced from the experimentally obtained variation of the averaged velocity of the fluid through the fabric with the prescribed difference between the upstream and downstream pressure values.

For each candidate equivalent pore shown in \Cref{fig: geometry}, seven pore-resolved, unsteady numerical simulations mimicking the experimental setup described above are performed using direct numerical simulation (DNS) flow solver Hydrodynamics Adaptive Mesh Refinement Simulator (HAMeRS) \cite{wong2019high} as described in \Cref{sec:DNS}. 
In all simulations, the computational fluid domain is chosen to be the $4\ \mathrm{mm} \times 0.5\ \mathrm{mm} \times 0.5\ \mathrm{mm}$ box (dimensions in the $x$-, $y$-, and $z$-directions, respectively)
shown in \Cref{fig: computational domain}. The upstream (inlet) boundary conditions as well as the initial conditions in the half domain delimited in the $x$-direction by the inlet 
boundary and the porous wall are set to the uniform flow field defined by $\rho^{\textrm{up}} = 1.111\ \mathrm{kg/m}^3$, ${\bm{v}}^{\textrm{up}} = 0\ \mathrm{m \ s^{-1}}$, 
and $p^{\textrm{up}} = 94,625\ \mathrm{Pa}$. As for the downstream (outlet) boundary conditions and the initial conditions in the other half of the computational fluid domain, they are  
set in each case to the uniform flow field defined by $\rho^{\textrm{dn}} = 1.112\ \mathrm{kg/m}^3$ ($1.109\ \mathrm{kg/m}^3$, 
$1.105\ \mathrm{kg/m}^3$, $1.099\ \mathrm{kg/m}^3$, $1.088\ \mathrm{kg/m}^3$, $1.076\ \mathrm{kg/m}^3$, and $1.064 \mathrm{kg/m}^3$), ${\bm{v}}^{\textrm{dn}} = 0\ \mathrm{m \ s^{-1}}$, and 
$p^{\textrm{dn}} = 94,525\ \mathrm{Pa}$ ($94,425\ \mathrm{Pa}$, $94,125 \ \mathrm{Pa}$, $93,625\ \mathrm{Pa}$, $92,625\ \mathrm{Pa}$, $91,625\ \mathrm{Pa}$, and 
$90,625 \mathrm{Pa}$). Note that these initial conditions are consistent with the temperature $T = 296.5 \ \mathrm{K}$ of the quiescent air in the laboratory.

\begin{figure}[!ht]
  \centering
  \includegraphics[width=1.0\linewidth]{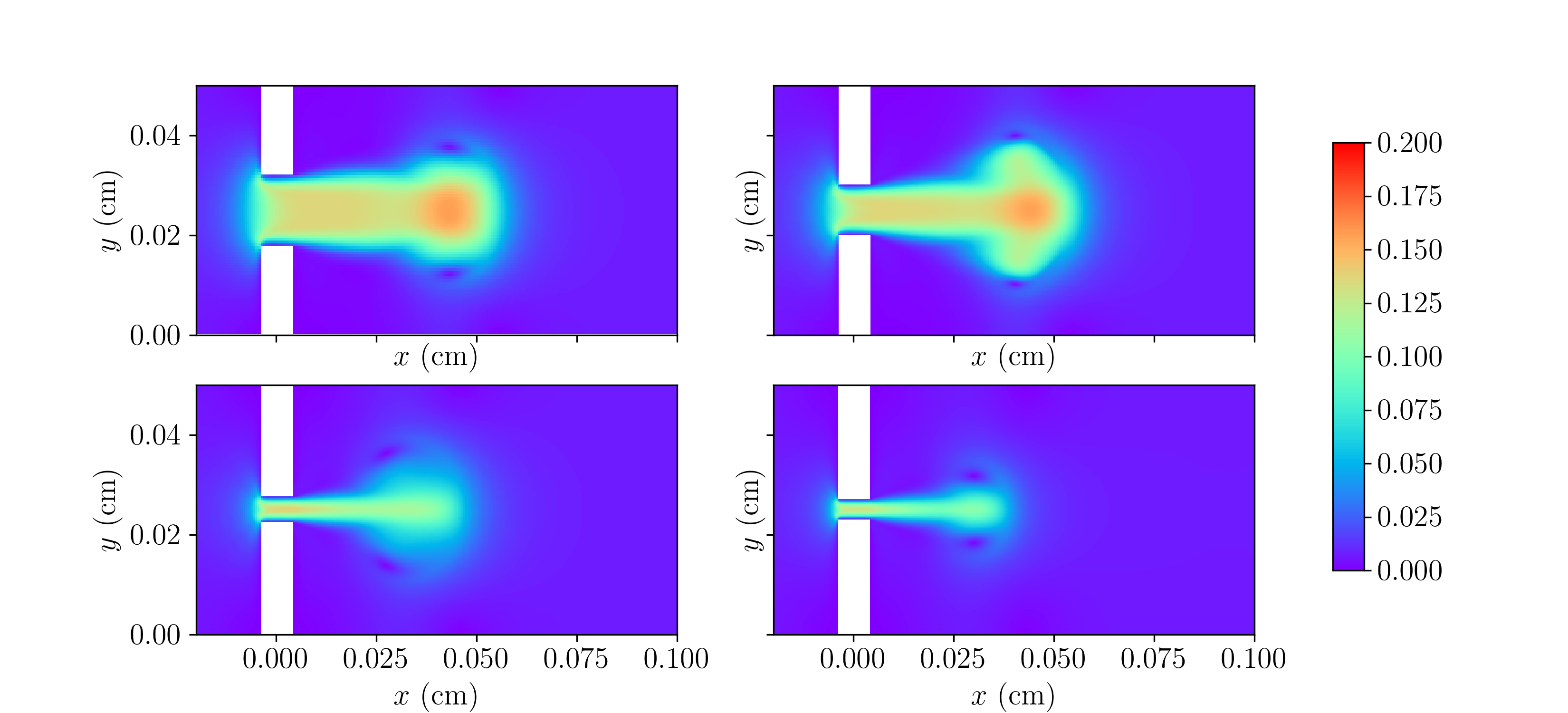}
  \caption{Mach number profiles in the central $x$-$y$ plane computed using pore-resolved numerical simulations and different pore geometries: $2/7 \times 2/7$ (top-left); 
	$20\% \times 40\%$ (top-right); $10\% \times 80\%$ (bottom-left); and $8\% \times 100\%$ (bottom-right) ($p^{up} - p^{dn} = 3,000 \ \mathrm{Pa}$, $t=15.8 \ \mathrm{\mu s}$).}
  \label{fig: NASA_Mach}
\end{figure}

At the beginning of each pore-resolved, unsteady numerical simulation characterized by a different uniform outlet pressure field, pressure waves propagate in both directions away from the porous fabric 
coupon. For each explored pore geometry, the pressure difference across the porous wall instantly drops, the fluid flow near the porous wall accelerates, and a weak jet stream forms downstream.
In \Cref{fig: NASA_Mach}, the reader can observe that the computed Mach number profiles substantially differ across the considered pore geometries. As expected, the holes with the smaller aspect ratios
($2/7\times2/7 \textrm{ and } 20\% \times 40\%$) lead to larger maximum Mach number values (around 0.2), which is consistent with the observation made in \Cref{sec:BFT} that equivalent 
pores with smaller aspect ratios lead to smaller momentum dissipation.

\begin{figure}[!ht]
\centering
\includegraphics[width=0.6\linewidth]{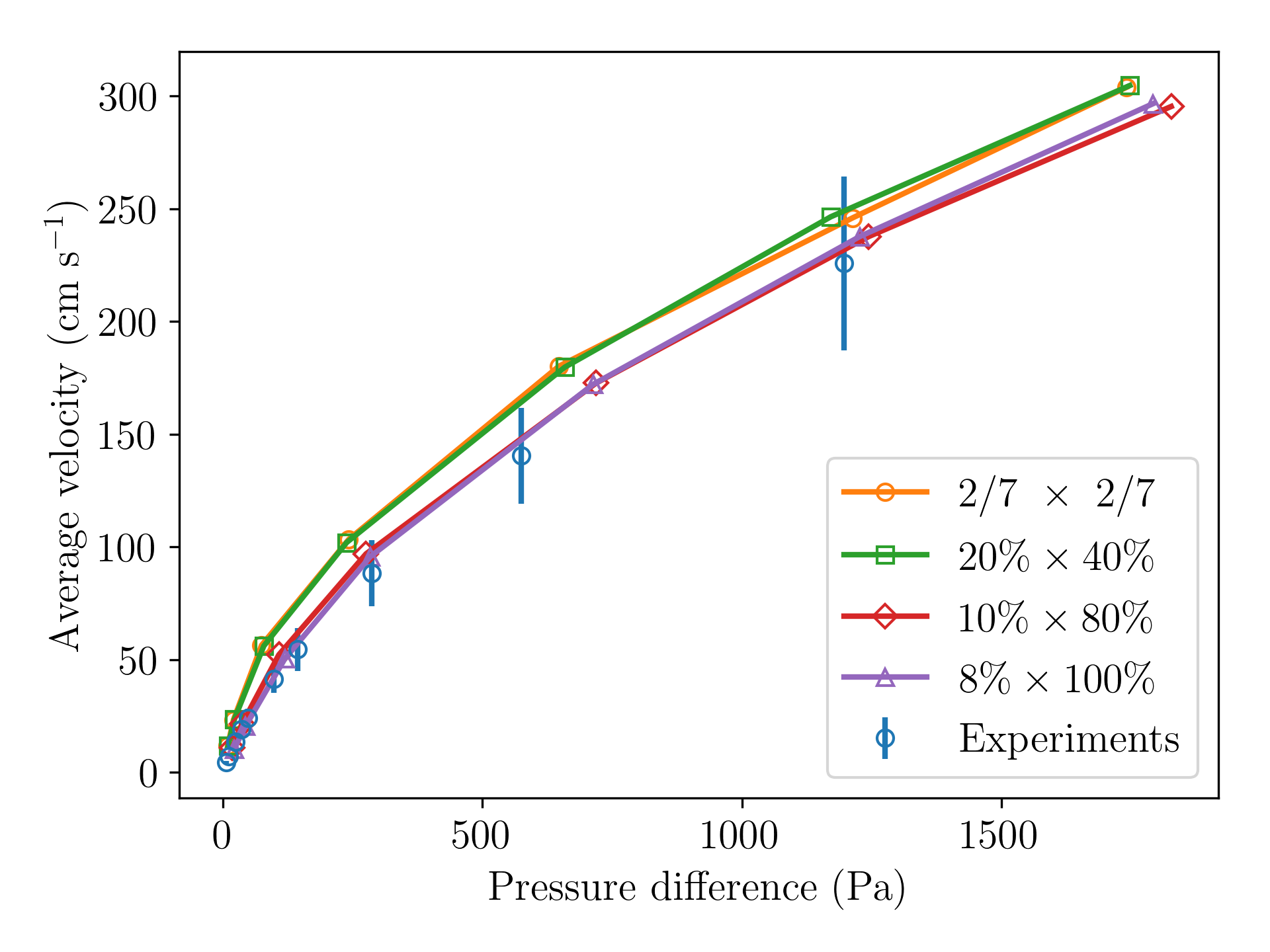}
	\caption{Permeability results (PIA-C-7020 Type I fabric with Ripstop weave): experimental data (blue circles and 99.7\% confidence interval); and quasi-steady results of pore-resolved,
	unsteady numerical simulations (other symbols).}
\label{fig: validation}
\end{figure}

In each of the 28 performed pore-resolved numerical simulations, the pressure difference across the porous wall rapidly tends to become quasi-steady and the velocity field on each side of the porous wall 
converges to a steady-state value around $t=15.8 \ \mathrm{\mu s}$. These results are reported in \Cref{fig: validation} in the form of velocity-pressure difference curves together with their counterpart 
experimental values measured in \cite{cruz2017permeability} within a confidence interval of $99.7\%$. Here, velocity refers to the averaged quasi-steady velocity through the fabric and pressure difference
refers to the averaged quasi-steady pressure difference between both sides of the porous wall.

The permeability velocity through the porous membrane $u^{\textrm{pe}}$ is defined as the ratio of the averaged mass flux across $\Omega$ and the initial upstream density 
$\rho^{\textrm{up}}$ -- that is,
\begin{equation}
	u^{\textrm{pe}} = \displaystyle{\frac{\int_{\Omega} \rho u \, d\Omega}{|\Omega| \rho^{\textrm{up}}}}\,.
	\label{eq: perm}
\end{equation}
Both experimental and numerical results reported above are close to the incompressible flow regime, where the density fluctuation can be neglected. Indeed, all performed pore-resolved numerical 
simulations reveal a density fluctuation of the order of $2\%$. In this case, the definition (\ref{eq: perm}) simplifies to $u^{\textrm{pe}} = \bar u$. From this result and
\Cref{fig: validation}, it follows that: each explored pore geometry leads to a numerical prediction of the permeability velocity that matches reasonably well its experimentally measured counterpart; 
but the geometry labeled as (d) in \Cref{fig: geometry}, which is characterized by the highest aspect ratio ($8\% \times 100\%$ or 12.5:1)
\footnote{The geometry labeled as (c) in \Cref{fig: geometry} delivers similar results because its aspect ratio is very close to that of the geometry labeled as (d).}, 
delivers the closest permeability velocity value to the experimentally measured one and therefore is the most suitable equivalent pore geometry.

It is worth noting that according to \cite{ergun1949fluid} and for a given wall permeability, the total pressure drop across a porous wall $(\Delta p)^{\textrm{tot}}$ 
can be split into two components due to viscous and inertial effects -- that is,
\begin{equation}
    \Delta p^{\textrm{tot}} = \Delta p^{\textrm{vis}} + \Delta p^{\textrm{ine}}.
	\label{eq:DPTOT}
\end{equation}
The viscous effect is dominant at low Reynolds number and the inertial effect prevails at high Reynolds number. The study carried out herein falls into the former category ($10 \le Re \le 400$), 
where one contribution of the viscous effect is the friction loss due to the shear stress near the surface of each spun fiber. Pore geometries characterized by smaller aspect ratios 
(e.g., $2/7 \times 2/7$ and $20\% \times
40\%$) are also characterized by smaller pore surfaces (e.g., see \Cref{fig: geometry}): hence, they lead to smaller values of $\Delta p^{\textrm{vis}}$ and therefore smaller values of
$\Delta p^{\textrm{tot}}$. This explains why the pore geometries with the larger aspect ratios (e.g., $10\% \times 80\%$ and $8\% \times 100\%$) are found in \Cref{fig: validation} to lead to numerical 
predictions of the permeability velocity that match better the experimental data.

\subsection{Verification for a supersonic application}
\label{sec:VSA}

Here, the homogenization approach presented in this paper is verified for a series of supersonic instances of the porous flow model problem graphically depicted in \Cref{fig: computational domain}.
In each instance: 
\begin{itemize} 
	\item The computational fluid domain is chosen to be the $10\ \mathrm{mm} \times 0.5\ \mathrm{mm} \times 0.5\ \mathrm{mm}$ box (dimensions in the $x$-, $y$-, and $z$-directions, respectively).
	\item This domain is divided in two non overlapping subdomains labeled as ``Upstream'' and ``Downstream'': the interface $\Gamma$ between them is located at $250 \ \mathrm{\mu m}$ from the 
		porous wall within the Upstream subdomain (see \Cref{fig: Mars initial flow} below).  
		\begin{figure}[!ht] 
			\centering 
			\centering 
			\begin{tikzpicture}[scale=2] 
				\draw[thick,->] (-0.9, 0.3, 0) -- (-0.4, 0.3, 0) node[anchor=north]{$x$}; 
				\draw[thick,->] (-0.9, 0.3, 0) -- (-0.9, 0.8, 0) node[anchor=south]{$y$};
				\draw[thick,->] (-0.9, 0.3, 0) -- (-0.9, 0.3, 0.5) node[xshift=-5][anchor=south]{$z$};
				\draw [fill=white,line width=0.2mm] (0,0) rectangle (5,1);
				\draw [black, thick, fill=cyan] (3 - 0.08,0) rectangle (3 + 0.08,1.0);
				\draw [red, line width=0.2mm] (2.6, 0) rectangle (2.6, 1.0);
				\draw node at (2.6, 1.2) {$\Gamma$};
				\draw node at (3.0, -0.2) {Porous wall};
				\draw node at (1.0, 0.5) {Upstream}; \draw node at (2.3, 0.5) {Shock}; \draw node at (4.0, 0.5) {Downstream}; 
				\draw node at (4.1, 0.5) {}; 
			\end{tikzpicture} 
			\caption{Supersonic porous flow model problem: separation of the computational fluid domain in two subdomains; interface $\Gamma$; and initial shock (central $x$-$y$ plane 
			view).} 
			\label{fig: Mars initial flow} 
		\end{figure}
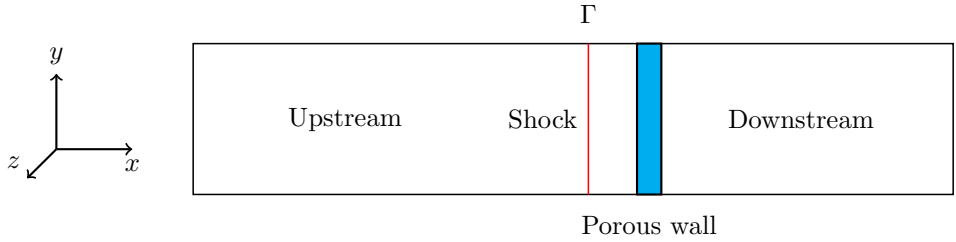
	\item In both subdomains, the flow solution is initialized using constant states chosen so that at $t = 0$, a shock occurs at $x = \Gamma$ with a specified initial speed $V_s$. Specifically:
		\begin{itemize}
			\item In the Upstream subdomain, the flow solution is initialized as follows
				\begin{eqnarray*} 
					\rho^{\textrm{up}} (x, y, z, 0) &=& 0.0076 \ \mathrm{kg \ m^{-3}}\,, \quad  p^{\textrm{up}} (x, y, z, 0) = 260\ \mathrm{Pa}\,,\\
					u^{\textrm{up}}(x, y, z, 0) &=& u^{\textrm{up}}(V_s, \textrm{RH})\,, \quad{\textrm{and}} \quad v(x, y, z, 0) = w(x, y, z, 0) = 0\,,
				\end{eqnarray*}
				where the notation $\spadesuit(V_s, \textrm{RH})$ designates that $\spadesuit$ is determined from $V_s$ and the Ranking-Hugoniot jump conditions.
			\item In the Downstream subdomain, the flow is assumed to be initially quiescent and in a constant state determined by $V_s$ and the Rankine-Hugoniot jump 
				conditions (see \Cref{tab: initial condition}) -- that is
				 \begin{eqnarray*} 
					 \rho^{\textrm{dn}} (x, y, z, 0) &=& \rho^{\textrm{dn}} (V_s, \textrm{RH})\,, \quad p^{\textrm{dn}} (x, y, z, 0) = p^{\textrm{dn}}(V_s, \textrm{RH})\,,\\
					 \textrm{and} \quad u^{\textrm{dn}}(x, y, z, 0) &=& v(x, y, z, 0) = w(x, y, z, 0) = 0.
				 \end{eqnarray*}
			\item The Mars atmosphere consisting mainly of carbon dioxide CO$_2$, the gas properties are set to
				\begin{equation}
					   \gamma=1.33, \quad R=188.4 \ \mathrm{J \ kg^{-1}K^{-1}},\quad \textrm{and} \quad \mu = 1.03\times10^{-5} \  \mathrm{kg \ m^{-1}s^{-1}}.
				\end{equation}
		\end{itemize}
	\item At each subdomain boundary facing $\Gamma$ -- that is, at each of the inlet and outlet boundaries -- the boundary conditions are set to the uniform flow defined by the constant state 
		used to specify the initial conditions in that subdomain.
	\item Two equivalent pore geometries defined by the aspect ratios $2/7\times2/7$ and $8\% \times 100\%$ ($\alpha = 8\%$) are considered.
	\item For each considered equivalent pore geometry, three unsteady numerical simulations are performed in a time-interval sufficiently large to capture the reflection of the shock wave on the 
		porous wall: 
		\begin{itemize}
			\item One three-dimensional pore-resolved simulation performed using DNS flow solver HAMeRS as described in \Cref{sec:DNS} to generate reference data.
			\item Two counterpart one-dimensional simulations performed using the flow solver AERO-F \cite{AS1, AS2} equipped with the EBM FIVER 
				\cite{wang2011algorithms, lakshminarayan2014embedded,main2017enhanced, huang2018family} and the homogenization approach presented in this paper 
				(see \Cref{sec: implementation}): one simulation performed using 400 grid points; and one performed using 2,000 grid points.
		\end{itemize}
\end{itemize}

Each aforementioned supersonic instance of the porous flow model problem graphically depicted in \Cref{fig: computational domain} is defined by a specified shock speed value $V_s$.
Three different values of $V_s$ are considered: $V_s = 279.4 \ \mathrm{m \ s^{-1}}$, $V_s = 335.9 \ \mathrm{m \ s^{-1}}$, and $V_s = 367.1 \ \mathrm{m \ s^{-1}}$. In all cases,
the effect of bulk viscosity is neglected.

\begin{table}[ht]
\begin{center}
\begin{tabular}{||c|c|c||} 
\hline
\hline
$V_s$ & Upstream boundary/initial conditions & Downstream boundary/initial conditions\\
\hline
$279.4 \ \mathrm{m \ s^{-1}}$ & $\rho=0.0076 \ \mathrm{kg \ m^{-3}}$, $u =127.98 \ \mathrm{m \ s^{-1}}$, $p = 260 \ \mathrm{Pa}$   & $\rho=0.004119 \ \mathrm{kg \ m^{-3}}$, $u =0 \ \mathrm{m \ s^{-1}}$,  $p = 112.73 \ \mathrm{Pa}$ \\ 
$335.9 \ \mathrm{m \ s^{-1}}$ & $\rho=0.0076 \ \mathrm{kg \ m^{-3}}$, $u =213.31 \ \mathrm{m \ s^{-1}}$, $p = 260 \ \mathrm{Pa}$   & $\rho=0.002774 \ \mathrm{kg \ m^{-3}}$, $u =0 \ \mathrm{m \ s^{-1}}$,  $p = 61.240 \ \mathrm{Pa}$ \\ 
$367.1 \ \mathrm{m \ s^{-1}}$ & $\rho=0.0076 \ \mathrm{kg \ m^{-3}}$, $u =255.97 \ \mathrm{m \ s^{-1}}$, $p = 260 \ \mathrm{Pa}$   & $\rho=0.002300 \ \mathrm{kg \ m^{-3}}$, $u =0 \ \mathrm{m \ s^{-1}}$,  $p = 43.772 \ \mathrm{Pa}$ \\ 
\hline
\hline
\end{tabular}
	\caption {Supersonic porous flow model problem: boundary and initial conditions (additionally, $v = w = 0 \ \mathrm{m \ s^{-1}}$ in all cases and both subdomains).}
\label{tab: initial condition}
\end{center}
\end{table}

\subsubsection{Pore-resolved numerical simulations}
\label{sec:PRNS}

\begin{figure}[!ht]
  \centering
  \includegraphics[width=1.0\linewidth]{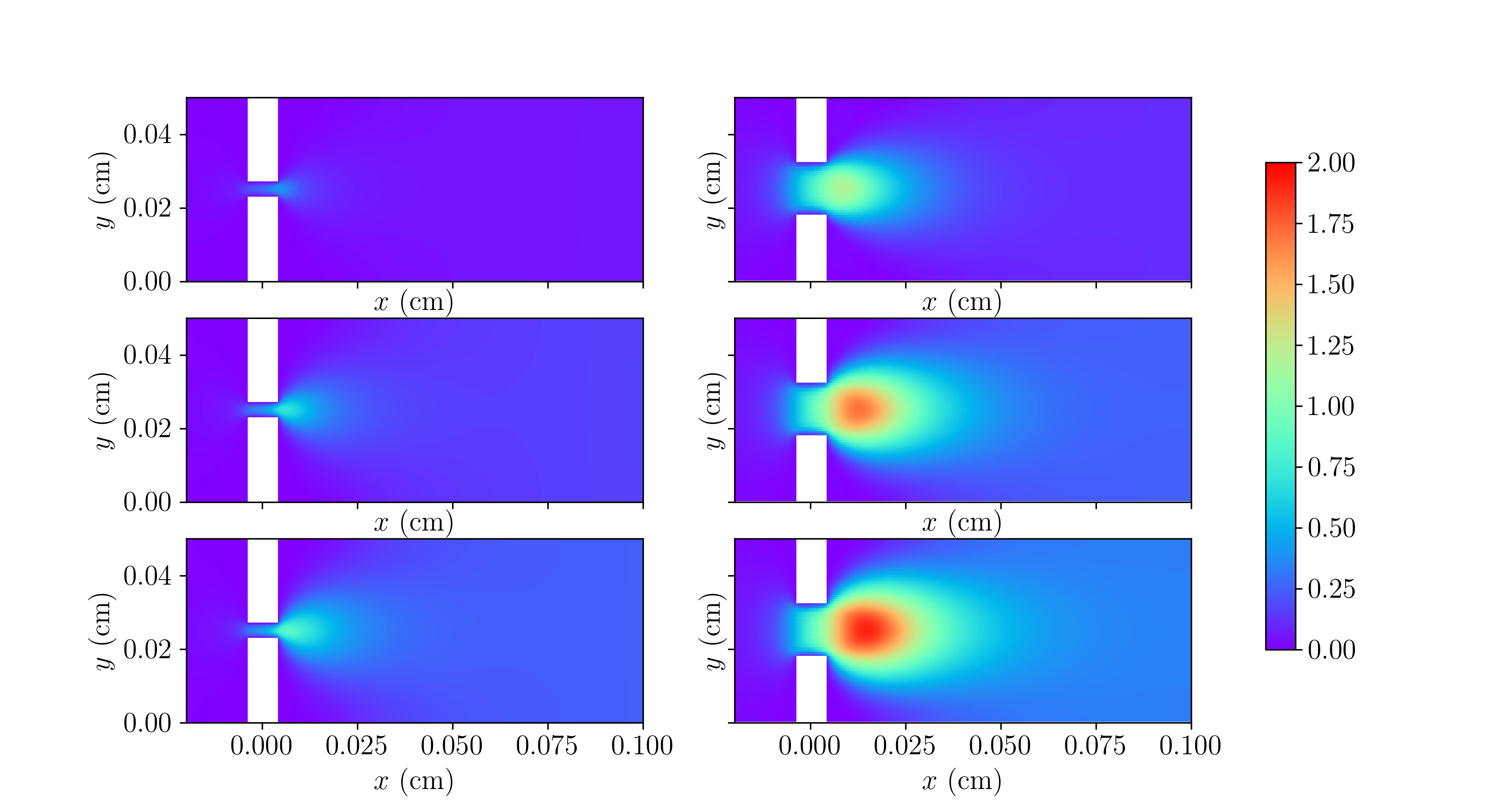} 
	\caption{Mach number profiles in the central $x$-$y$ plane computed at $t = 23.46 \ \mathrm{\mu s}$ using pore-resolved, unsteady numerical simulations: pore geometry with $2/7 \times 2/7$ 
	aspect ratio (left) and $8/\% \times 100/\%$ aspect ratio (right); $V_s = 279.4, \ 335.9, \ \textrm{and} \ 367.1 \ \mathrm{m \ s^{-1}}$ (top to bottom).}
  \label{fig: Mars_Mach}
\end{figure}

For each considered pore geometry and each considered shock speed value $V_s$, the results of the pore-resolved, unsteady simulation reveal that when the shock bounces off the porous wall, a high 
density/high pressure flow zone develops upstream of the porous wall. Consequently, the flow through the pore behaves like a flow through a converging-diverging nozzle. Downstream of the porous wall, the 
flow immediately accelerates and goes through an expansion. \Cref{fig: Mars_Mach} displays the Mach number profiles computed at $t = 23.46 \ \mathrm{\mu s}$. It shows that as in the case of low-speed 
flow (see \Cref{fig: NASA_Mach}), the pore geometry with the smaller aspect ratio leads to jet streams. However, the reader can also observe that in the high-speed case, the flows are choked.

Let 
\begin{equation}
	Re = \displaystyle{\left( \frac{\int_{\Omega^{\textrm{pore}}} \rho u^2 \, d\Omega^{\textrm{pore}}}{\rho^{\textrm{up}} |\Omega^{\textrm{pore}}|} \right)^{1/2} \frac{\rho^{\textrm{up}} d}{\mu}},
\end{equation}
where $|\Omega^{\textrm{pore}}|$ denotes the area of the pore and $d = \sqrt{4|\Omega^{\textrm{pore}}|/\pi}$ is the pore diameter, denote the Reynolds number based on the jet flow velocity at the pore 
exit. \Cref{tab: Reynolds number} reports the values of this number computed at $t = 23.46 \ \mathrm{\mu s}$ 
by post-processing the results of the pore-resolved simulations. The reported low values, which are partially due to the low 
density and partially to the small pore size, highlight the importance of accounting for viscous effects when computing flows past porous walls

\begin{table}[ht]
\begin{center}
\begin{tabular}{ c|c c c } 
\hline
\hline
Aspect ratio of pore geometry& $V_s = 279.4 \ \mathrm{m \ s^{-1}}$ & $V_s = 335.9 \ \mathrm{m \ s^{-1}} $ & $V_s = 367.1 \ \mathrm{m \ s^{-1}} $\\
\hline
$2/7 \times 2/7$   &  7.3  & 13.50 & 17.96 \\
$8\% \times 100\%$ & 16.43 & 23.75 & 27.38\\ 
\hline
\hline
\end{tabular}
	\caption {Reynolds numbers based on the jet flow velocity computed at $t = 23.46 \ \mathrm{\mu s}$ using pore-resolved numerical simulations.}
\label{tab: Reynolds number}
\end{center}
\end{table}

\begin{figure}[!ht]
  \centering
  \includegraphics[width=0.6\linewidth]{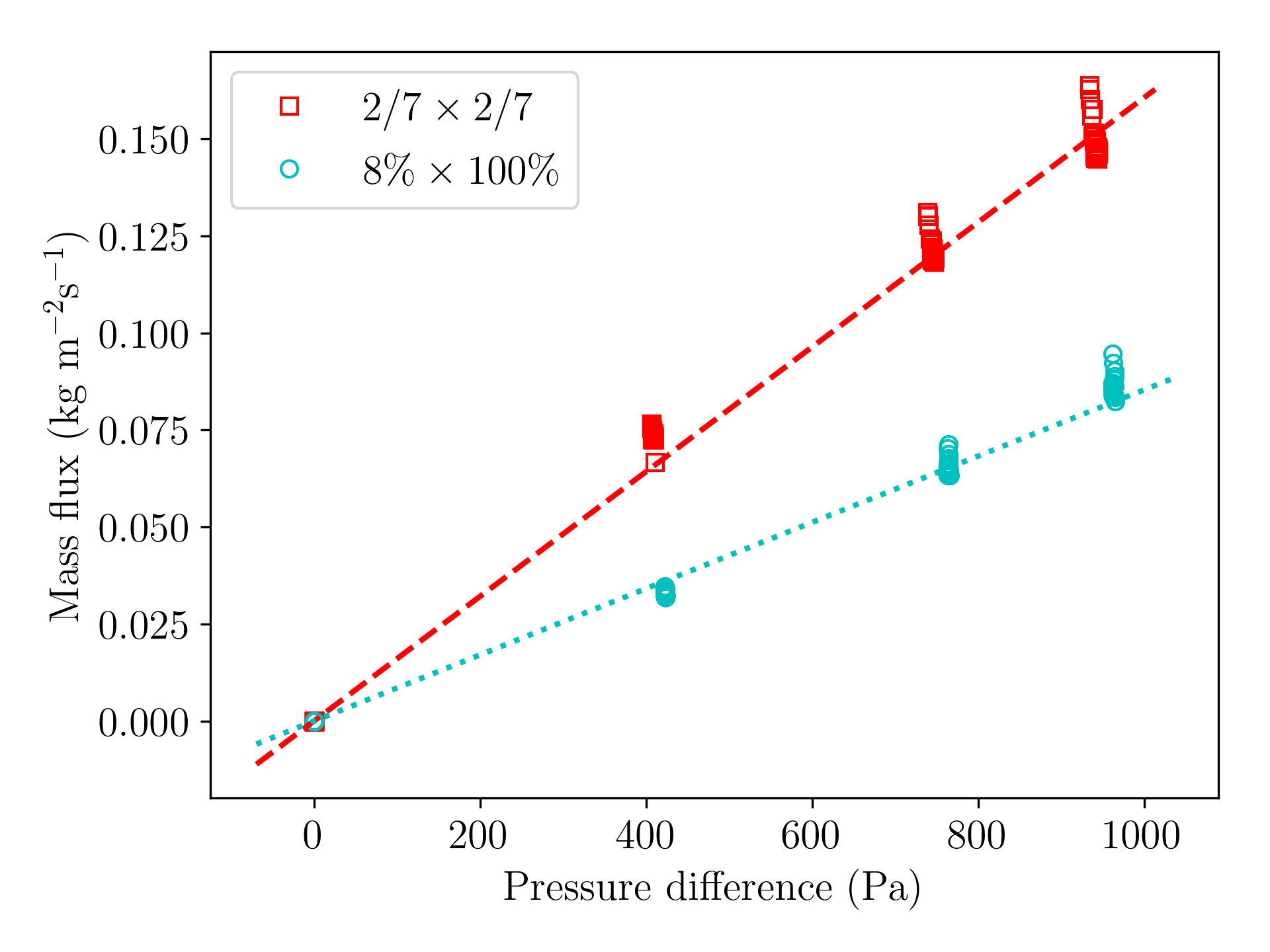}
  \caption{Variations of the averaged mass flow rate with the pressure difference computed using pore-resolved, unsteady numerical simulations.}
  \label{fig: Pressure_Massflux_Relation}
\end{figure}

\Cref{fig: Pressure_Massflux_Relation} reports for each considered pore geometry the variation of the averaged mass flow rate with the pressure difference between both sides of the 
porous wall computed by post-processing at each time-step of the computation the results of the corresponding pore-resolved, unsteady numerical simulation. Each variation is found to be linear 
and therefore describable by
\begin{equation}
    \overline{\rho u} \approx K \overline{\Delta p}.
\end{equation}
In other words, it satisfies Darcy's law, where the constant $K$ depends on the pore geometry. In this case, a simple least square fitting yields $K = 1.61 \times 10^{-4} \ \mathrm{s \ m^{-1}}$ for the 
pore geometry with $2/7 \times 2/7$ aspect ratio and $K = 8.54\times 10^{-5} \ \mathrm{s \ m^{-1}}$ for that with $8\% \times 100\%$ aspect ratio.

\subsubsection{Homogenization-based numerical simulations}
\label{sec:HBNS}

Recall that all numerical simulations discussed herein are performed at the coupon level, due to the periodicity of the weave pattern (see \Cref{sec:SETUP}). For this reason and due to the 
flat geometry of the considered porous wall (see \Cref{fig: Mars initial flow}), all counterparts of the pore-resolved, unsteady numerical simulations presented in \Cref{sec:PRNS} based on
the homogenization approach described in this paper are performed in one dimension, in the computational fluid domain $x \in [-0.5, 0.5] \ \mathrm{cm}$. Two discretizations of this domain in
dual computational cells are considered for this purpose: one with $N = 400$ grid points; and one with $N = 2,000$ grid points. In each case, the porous wall is ``embedded'' in the discretization
at $x = 0$. The homogenization-based numerical simulations are performed using the vertex-based, finite volume flow solver AERO-F~\cite{AS1, AS2} equipped with the second-order 
space-accurate EBM FIVER \cite{main2017enhanced} (see also \Cref{sec: implementation}) and the second-order explicit Runge-Kutta time-integrator constrained by the constant CFL number of 0.5. 

Because for the equivalent pore with the aspect ratio of $2/7 \times 2/7$ the jet stream was reported in \Cref{sec:PRNS} to expand immediately downstream of the porous wall (see \Cref{fig: Mars_Mach}), 
the thickness correction factor (see \Cref{sec:BFT}) is set for this case to $\eta_f = 1$. For the equivalent pore with the aspect ratio of $8\% \times 100\%$, the jet stream can be seen in
\Cref{fig: Mars_Mach} to extend further in the computational fluid domain: thus, the thickness correction factor is set in this case to $\eta_f = 4$. 

\begin{figure}[!ht]
  \centering
  \includegraphics[width=0.4975\linewidth]{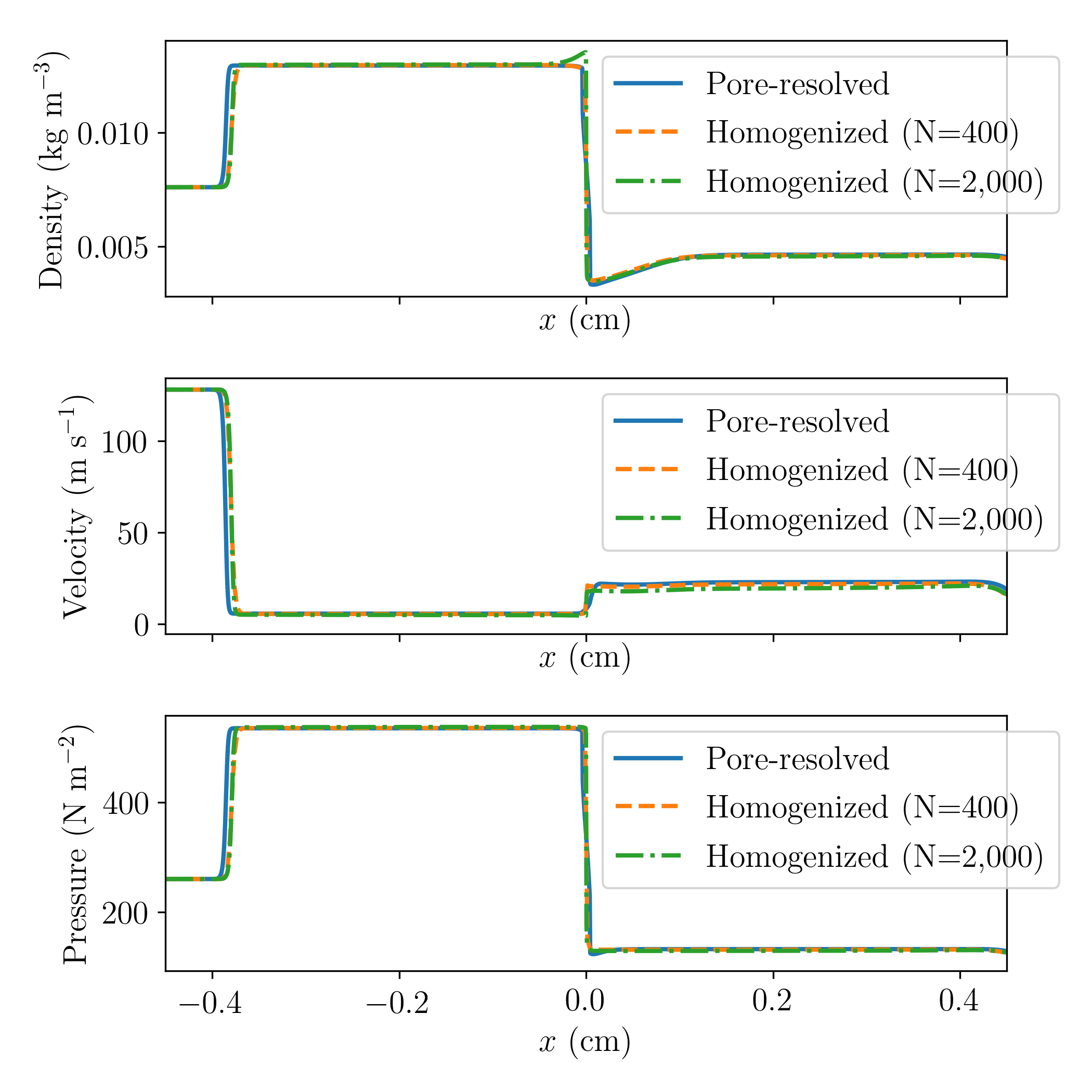}
  \includegraphics[width=0.4975\linewidth]{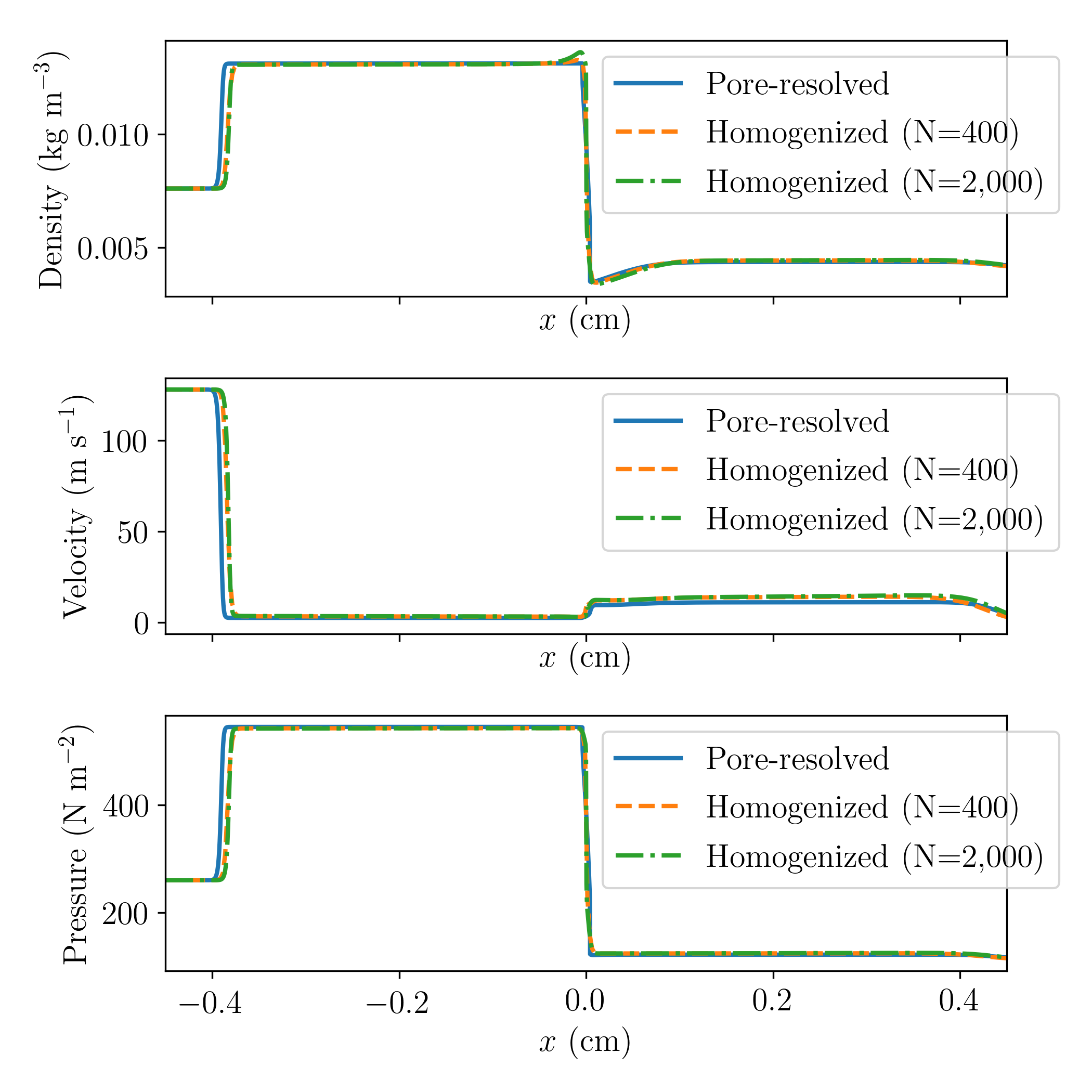}
	\caption{Averaged density, streamwise velocity, and pressure profiles (top to bottom) predicted at $t = 23.46 \ \mathrm{\mu s}$ for $V_s = 279.4 \mathrm{m \ s^{-1}}$: pore with aspect ratio of 
	$2/7 \times 2/7$ (left); and pore with aspect ratio of $8\% \times 100\%$ (right).}
  \label{fig: 3d_M06 results}
\end{figure}

\begin{figure}[!ht]
  \centering
  \includegraphics[width=0.4975\linewidth]{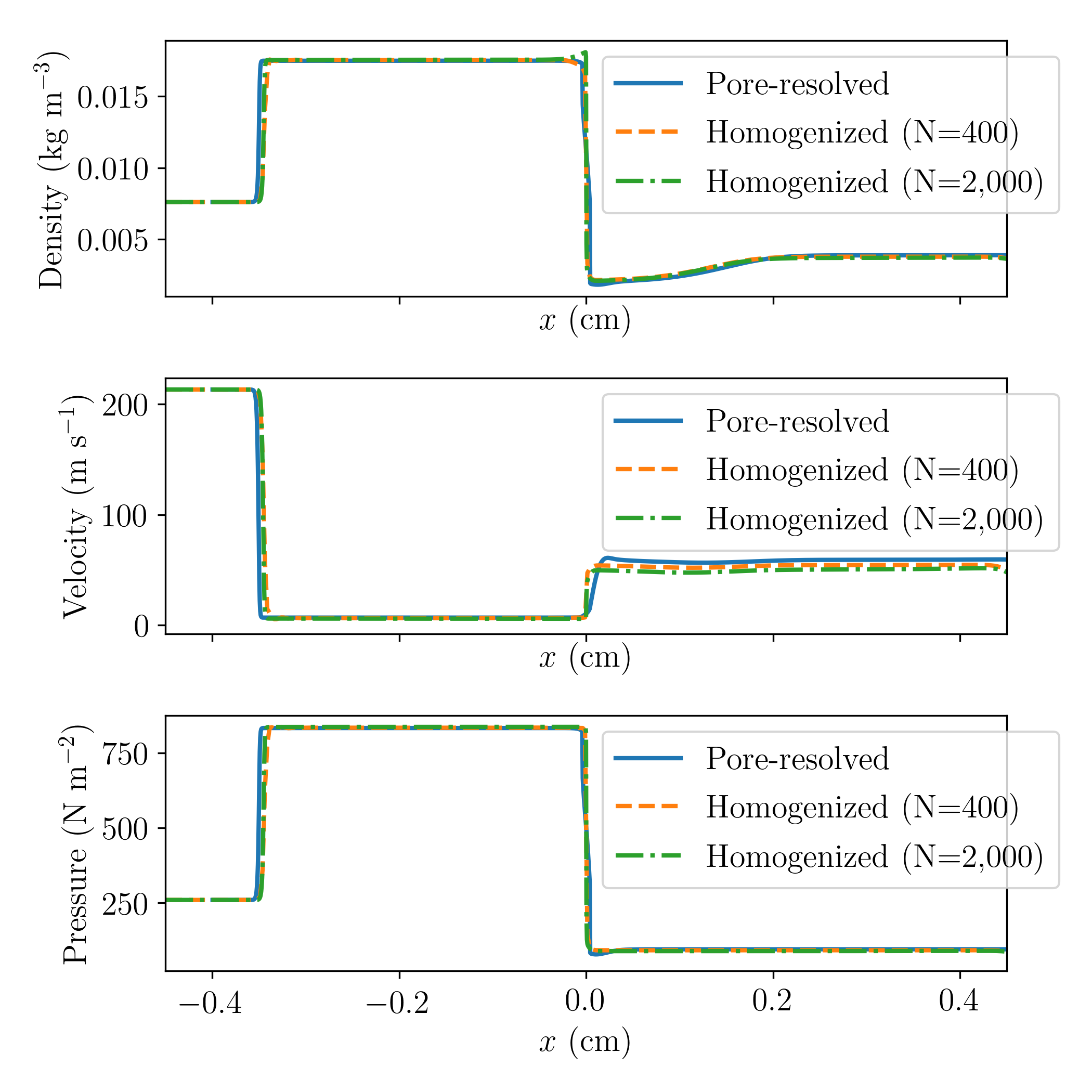}
  \includegraphics[width=0.4975\linewidth]{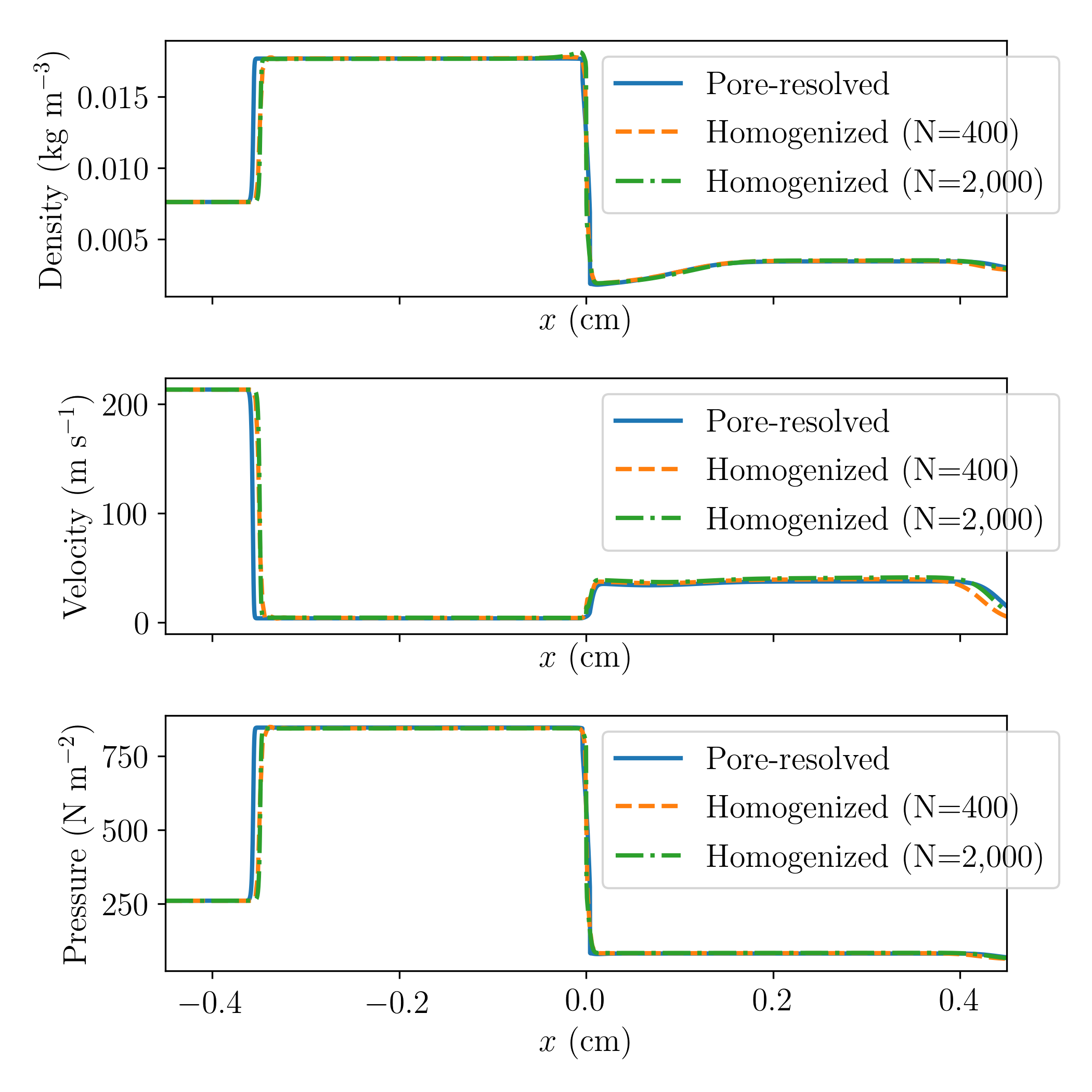}
      \caption{Averaged density, streamwise velocity, and pressure profiles (top to bottom) predicted at $t = 23.46 \ \mathrm{\mu s}$ for $V_s = 335.9 \mathrm{m \ s^{-1}}$: pore with aspect ratio of 
	$2/7 \times 2/7$ (left); and pore with aspect ratio of $8\% \times 100\%$ (right).}
  \label{fig: 3d_M1 results}
\end{figure}

\begin{figure}[!ht]
  \centering
  \includegraphics[width=0.4975\linewidth]{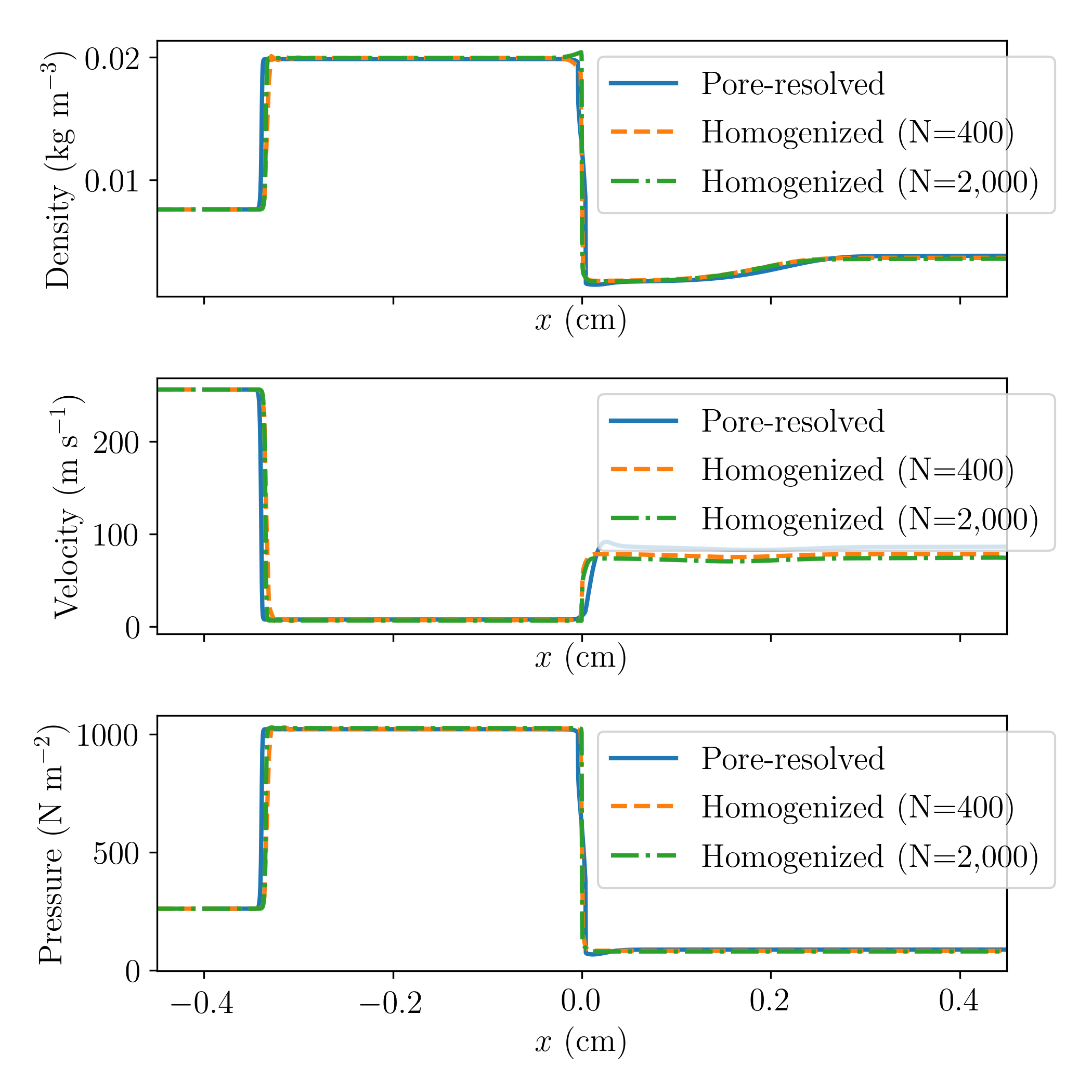}
  \includegraphics[width=0.4975\linewidth]{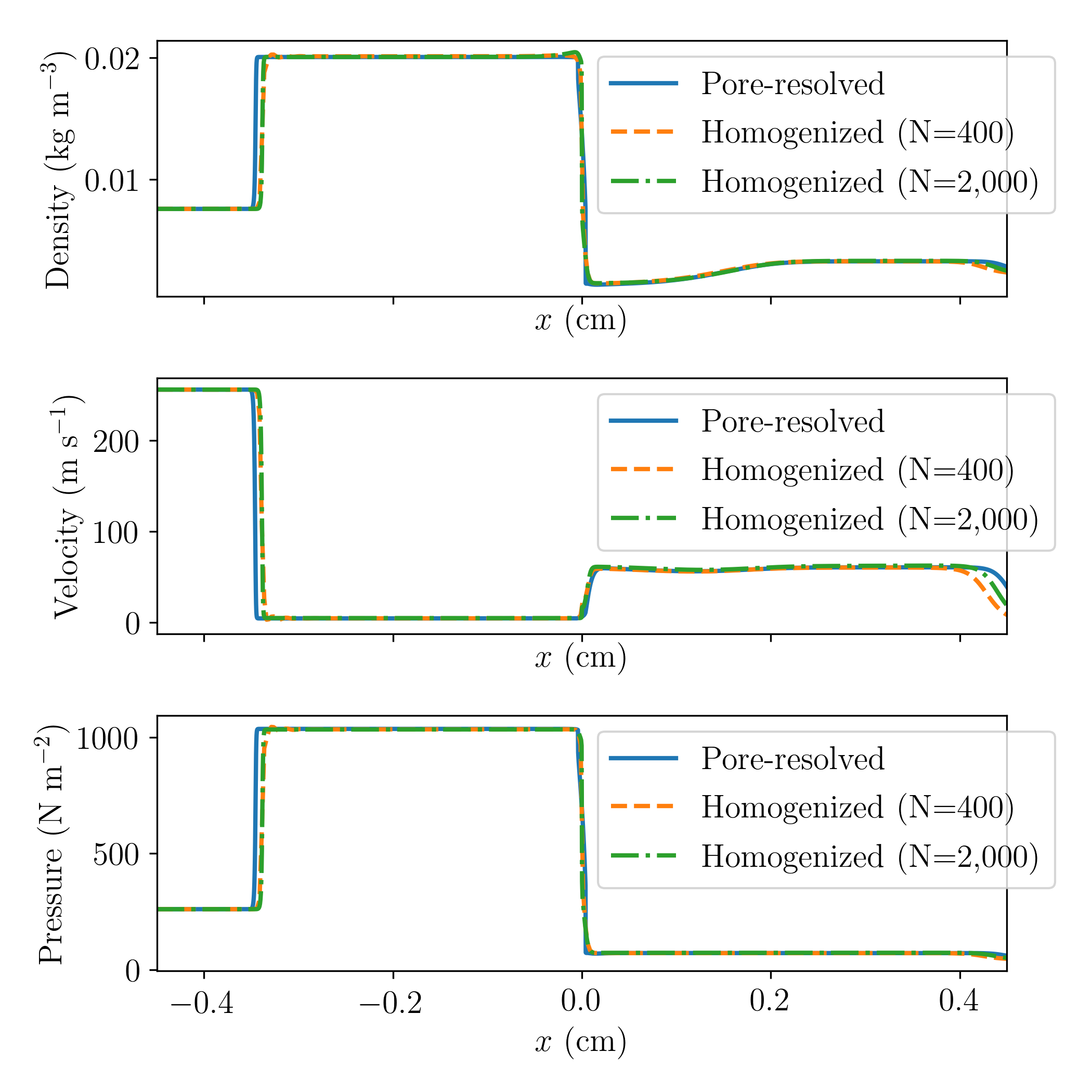} 
	\caption{Averaged density, streamwise velocity, and pressure profiles (top to bottom) predicted at $t=23.46 \ \mathrm{\mu s}$ for $V_s = 367.1 \mathrm{m \ s^{-1}}$: pore with aspect ratio of 
	$2/7 \times 2/7$ (left); and pore with aspect ratio of $8\% \times 100\%$ (right).}
  \label{fig: 3d_M12 results}
\end{figure}

\Cref{fig: 3d_M06 results,fig: 3d_M1 results,fig: 3d_M12 results} report the averaged density, streamwise velocity, and pressure profiles computed at $t=23.46 \ \mathrm{\mu s}$ 
using the proposed homogenization approach. It shows that: the one-dimensional discretization of $[-0.5, 0.5] \ \mathrm{cm}$ in $N = 400$ grid points is well resolved; and the results obtained
using the proposed homogenization approach agree well with their counterparts obtained using the pore-resolved numerical simulations. 

\newpage

Next, attention is focused on the drag force $F_D$ generated by the porous wall. For a pore-resolved numerical simulation, $F_D$ can be defined as
\begin{align}
\label{eq:drag-pore-resolved} 
	F_D =  \underbrace{- \int_{\partial{\Omega_{\textrm{vol}}}\backslash\partial\Omega_{\textrm{vol}}^{\textrm{pore}}} pn\  d\partial \Omega_{\textrm{vol}}}_{<1>}
	+      \underbrace{  \int_{\partial{\Omega_{\textrm{vol}}}\backslash\partial\Omega_{\textrm{vol}}^{\textrm{pore}}} \uptau \cdot n\ d\partial \Omega_{\textrm{vol}}}_{<2>}
	+      \underbrace{  \int_{\partial \Omega_{\textrm{vol}}^{\textrm{pore}}} \uptau \cdot n\ d\partial \Omega_{\textrm{vol}}^{\textrm{pore}}}_{<3>},
\end{align}
where $\Omega_{\textrm{vol}}$ denotes the volume representation of the porous wall, $\Omega_{\textrm{vol}}^{\textrm{pore}}$ denotes the part of $\Omega_{\textrm{vol}}$ occupied by the pores,
$\partial{\Omega_{\textrm{vol}}}$ denotes the {\it wet} boundary surface of $\Omega_{\textrm{vol}}$, $\partial \Omega_{\textrm{vol}}^{\textrm{pore}}$ denotes the boundary surface of
$\Omega_{\textrm{vol}}^{\textrm{pore}}$, and $n$ denotes the unit, outward normal to the surface over which the spatial integration is performed. Hence, $F_D$ consists in this case of a force due to 
the pressure jump across the porous wall (term <1> in (\ref{eq:drag-pore-resolved})), a force induced by the jump of the normal component of the shear stress traction across the porous wall (term <2> 
in (\ref{eq:drag-pore-resolved})), and the shear force on the pore boundaries (term <3> in (\ref{eq:drag-pore-resolved})).

For a simulation based on the homogenization approach proposed in this paper, $F_D$ can be defined consistently with (\ref{eq:drag-pore-resolved}) as follows
\begin{align}
\label{eq:drag-homogenized-general}
F_D =  -\int_{\Omega} p n\ d\partial \Omega + \int_{\Omega} \tau \cdot n\ d\partial \Omega\,,
\end{align}
which in the case of the one-dimensional homogenized computations performed herein simplifies to

\begin{align} 
	\label{eq:drag-homogenized-1D} 
	F_D =  -\int_{\Omega} p n\ d\partial \Omega + \int_{\Omega} \tau \cdot n\ d\partial \Omega  = |\Omega|(\Delta p^\textrm{tot}  - \Delta {\uptau}_{xx}) |\Omega|.
\end{align}

Each of the integrals in (\ref{eq:drag-pore-resolved}) and (\ref{eq:drag-homogenized-1D}) can be evaluated numerically using standard approximations. In the context of an EBM for CFD,
the load evaluation algorithm presented in \cite{lakshminarayan2014embedded} can be used for this purpose.

\begin{figure}[!ht]
  \centering
  \includegraphics[width=0.8\linewidth]{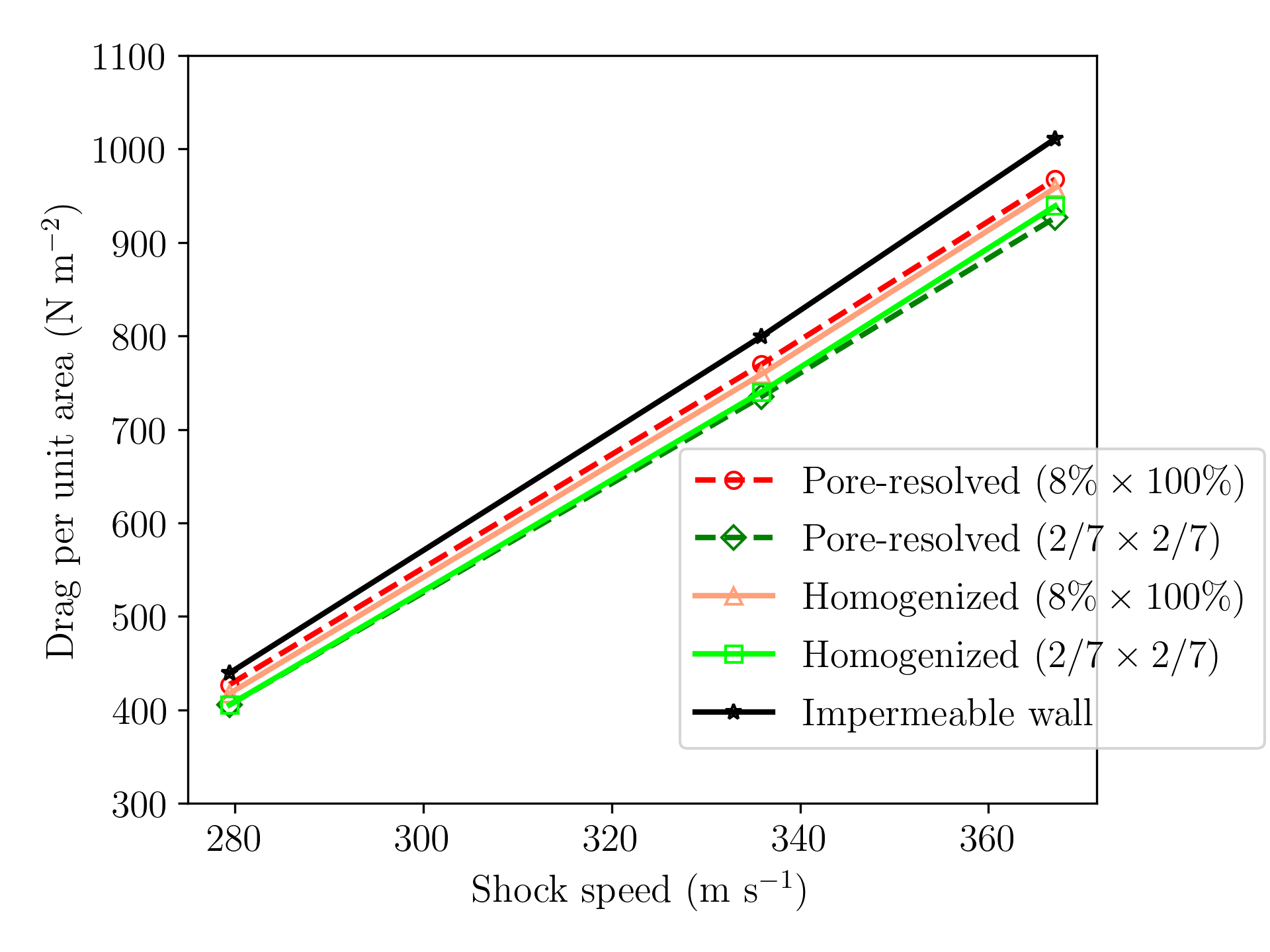}
	\caption{Variations with the shock speed of the drag force per unit area predicted at $t=23.46 \ \mathrm{\mu s}$ (for homogenization-based numerical simulations, $N = 400$).}
  \label{fig: drag results}
\end{figure}

\Cref{fig: drag results} shows that for $N = 400$, the homogenization-based numerical simulations accurately reproduce in all cases the variation with the shock speed of the drag force per unit area 
obtained at $t=23.46 \ \mathrm{\mu s}$ using the counterpart pore-resolved numerical simulation.

\section{Conclusions}\label{sec: conclusion}

The homogenization approach proposed in this paper for the treatment of porous wall boundary conditions distinguishes itself from alternatives in many ways. To begin, it is valid for both 
incompressible and compressible viscous flows. It does not require prescribing neither a mass flow rate nor a heuristic discharge coefficient. Instead, it models the inviscid flux through a porous
wall surrounded by the flow as a weighted average of the inviscid flux at an impermeable surface and that through pores, and introduces a body force term in the governing equations to account
for friction loss along the pore boundaries. The latter term is parameterized however by a thickness correction factor. The proposed approach,  which can be implemented in any 
semi-discretization or discretization method, is successfully verified for a series of one-dimensional supersonic flow problems through a rigid porous wall,
using three-dimensional pore-resolved numerical simulations.  Its successful application to three-dimensional fluid-structure interaction problems associated with supersonic
parachute inflation dynamics during Mars landing is reported in the companion papers ~\cite{huang2020modeling, avery2020computationally}.
\begin{appendices}
\label{appendix}

\crefalias{section}{appsec}

\section{EBM FIVER and arbitrary embedding CFD meshes}\label{sec: implementation}

Brief descriptions of the simplest version of the FInite Volume method with Exact two-material Riemann problems (FIVER) \cite{wang2011algorithms} and the implementation in this Embedded Boundary Method 
(EBM) for CFD of the proposed homogenization approach for the treatment of porous wall boundary conditions are provided here, in the context of arbitrary embedding CFD meshes. For more advanced versions 
of FIVER, the reader is referred to, for example, \cite{main2017enhanced, huang2018family, borker2019mesh}.

FIVER semi-discretizes the inviscid fluxes of \Cref{eq: ns} by a vertex-based Finite Volume (FV) method and the viscous fluxes of this equation by a FE method. Both methods are known for their
ability to operate on arbitrary meshes. Using the standard characteristic function associated with a control volume (dual cell) $\mathcal{C}_i$ defined at a grid point $i$ of the 
structured or arbitrarily unstructured CFD mesh, the standard piecewise-linear tent test function $\psi_i$ associated with the grid point $i$, the standard bijection between the space of piecewise 
linear FE functions defined over the primal elements of the CFD mesh and the space of constant characteristic shape functions defined over the dual cells of this mesh \cite{farhat1993two}, FIVER adopts 
the standard variational approach and integration by parts to semi-discretize \Cref{eq: ns} as follows
\begin{equation}\label{eq: ns semi-discrete}
\displaystyle{|\mathcal{C}_i|\dfrac{\partial \bm{W}_i}{\partial t}} +
\sum_{j \in \mathcal{K}(i)} 
\int_{\partial\mathcal{C}_{ij}}\mathcal{F}(\bm{W}_h) \boldsymbol \cdot n_{ij} \,d\partial B\, +
\int_{\partial\mathcal{C}_{i} \cap \partial B_\infty} \mathcal{F}(\bm{W}_h) \boldsymbol \cdot n_\infty \,d\partial B\, +
\sum_{ B^e \ni \, i}\int_{ B^e} \nabla{\psi^e_i} \boldsymbol \cdot \mathcal{G}(\bm{W}_h) \,d B^e\, = 0.
\end{equation}
In \Cref{eq: ns semi-discrete} above, $|\mathcal{C}_i|$ denotes the volume of $\mathcal{C}_i$, $\partial \mathcal{C}_i$ denotes its boundary, $\partial\mathcal{C}_{ij}$ denotes the cell facet shared by 
$\mathcal{C}_i$ and $\mathcal{C}_j$, $n_{ij}$ denotes the unit outward normal to $\partial \mathcal{C}_{ij}$, $\bm{W}_h$ denotes the spatial approximation of the fluid state vector $\bm{W}$, $\bm{W}_i$ 
denotes the average value of $\bm{W}_h$ in $\mathcal{C}_i$, $\mathcal{K}(i)$ denotes the set of grid points connected by an edge to the grid point $i$, $B$ denotes the computational fluid domain and 
$\partial B_\infty$ denotes its far-field boundary, $n_\infty$ denotes the unit outward normal to  $\partial B_\infty$, $B^e$ denotes a primal element of the CFD mesh discretizing $B$, and $\psi^e_i$ 
denotes the restriction of $\psi_i$ to $B^e$.

\begin{figure}[!ht] 
	\centering 
	\includegraphics[width=0.6\linewidth]{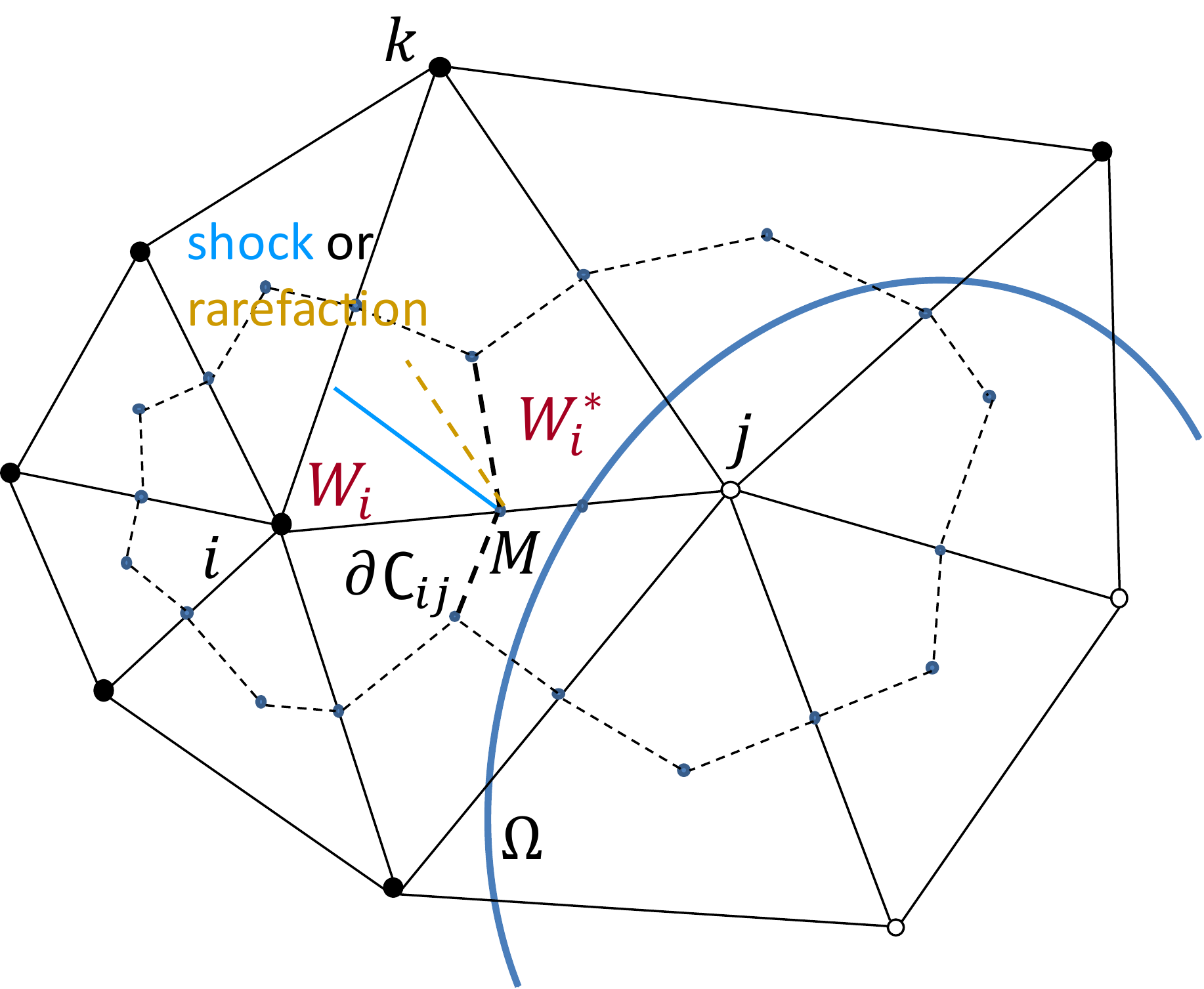} 
	\caption{Embedding of a porous wall $\Omega$ in a CFD mesh: construction and solution of a fluid-structure half-Riemann problem at the midpoint $M$ of the edge $(i, j)$ intersecting $\Omega$.} 
	\label{FIG: FIVER} 
\end{figure}

Let $\Omega$ denote a surface embedded in the discretization of $B$ by a CFD mesh and let $(i, j)$ be an edge of this mesh that intersects $\Omega$ (see \Cref{FIG: FIVER}). In order to 
semi-discretize the inviscid flux through $\partial\mathcal{C}_{ij}$, the simplest version of FIVER assumes that the midpoint $M$ of the edge $(i, j)$ lies on $\Omega$ -- and more specifically, 
$\partial\mathcal{C}_{ij}$ lies on $\Omega$; constructs a fluid-structure half-Riemann problem at the point $M$ in the direction of the normal $n_{ij}$ to $\partial\mathcal{C}_{ij}$; solves this 
half-Riemann problem to obtain the fluid state vector at the point $M$, $\bm{W}_{i}^\star$; and then approximates the inviscid flux through $\partial\mathcal{C}_{ij}$ as follows
\begin{eqnarray}
	\int_{\partial\mathcal{C}_{ij}}\mathcal{F}(\bm{W}_h) \boldsymbol \cdot n_{ij} \,dB
	&\approx& \bm{F}_{ij} = |\partial\mathcal{C}_{ij}| \, \bm{F}^{\textrm{\textrm{Roe}}}(\bm{W}_{i}, \bm{W}_{i}^\star, n_{ij}), \quad \textrm{if the grid point} ~ i ~ \textrm{lies in} ~ B \nonumber\\
	\hbox{and} && \label{eq: num flux fs}\\
	\int_{\partial\mathcal{C}_{ji}}\mathcal{F}(\bm{W}_h) \boldsymbol \cdot n_{ji} \,dB 
		&\approx& \bm{F}_{ji} = |\partial\mathcal{C}_{ji}| \, \bm{F}^{\textrm{\textrm{Roe}}}(\bm{W}_{j}, \bm{W}_{j}^\star, n_{ji}), \quad \textrm{if the grid point} ~ j ~ \textrm{lies in} ~ B, 
		\nonumber 
\end{eqnarray}
where $|\partial\mathcal{C}_{ij}|$ = $|\partial\mathcal{C}_{ji}|$ and $n_{ji} = -n_{ij}$. Note that by computing $\bm{W}_{i}^\star$ (or $\bm{W}_{j}^\star$) via the solution of a fluid-structure 
half-Riemann problem, FIVER is able to capture any shock or rarefaction wave near the embedded surface -- that is, at the fluid-structure interface. 
Away from $\Omega$ -- that is, when the edge $(i, j)$ does not intersect $\Omega$, FIVER approximates the inviscid flux through $\partial\mathcal{C}_{ij}$ using a standard FV method based 
on a standard approximate Riemann solver.

From (\ref{eq: num flux fs}) and \Cref{eq: homogeneous inviscid flux}, it follows that if $\Omega$ is a porous wall, the homogenized, numerical inviscid flux functions through $\Omega$ can be computed
by FIVER as follows
\begin{equation} 
	\begin{split} 
		\bm{F}^{\textrm{ave}}_{ij} &= |\partial{\mathcal C}_{ij}|\left(\alpha \bm{F}^{\textrm{\textrm{Roe}}}(\overline{\bm{W}}_i,\overline{\bm{W}}_j, n_{ij}) + (1 - \alpha) \bm{F}^{\textrm{imp-wall}}(\overline{\bm{W}}_i, n_{ij})\right) = 
		|\partial{\mathcal C}_{ij}|\left(\alpha \bm{F}^{\textrm{\textrm{Roe}}}(\overline{\bm{W}}_i,\overline{\bm{W}}_j, n_{ij}) + (1 - \alpha) \bm{F}^{\textrm{\textrm{Roe}}}(\bm{W}_{i}, \bm{W}_{i}^\star, n_{ij})\right)\,,\\ 
		\bm{F}^{\textrm{ave}}_{ji} &= |\partial{\mathcal C}_{ji}|\left(\alpha \bm{F}^{\textrm{\textrm{Roe}}}(\overline{\bm{W}}_j,\overline{\bm{W}}_i, n_{ji}) + (1 - \alpha) \bm{F}^{\textrm{imp-wall}}(\overline{\bm{W}}_j, n_{ji})\right) = |\partial{\mathcal C}_{ji}|\left(\alpha \bm{F}^{\textrm{\textrm{Roe}}}(\overline{\bm{W}}_j,\overline{\bm{W}}_i, n_{ji}) + (1 - \alpha) \bm{F}^{\textrm{\textrm{Roe}}}(\bm{W}_{j}, \bm{W}_{j}^\star, n_{ji})\right)\,,
	\end{split} 
\label{eq: FIVER homogeneous inviscid flux}
\end{equation}
where averaging is performed as in \eqref{eq:averaging}. Again, the second-order extension of the above approximations using the classical Monotonic Upstream-centered Scheme for Conservation Laws (MUSCL)
is straightforward.

As for the viscous approximation in \Cref{eq: ns semi-discrete}, it can be re-written as
\begin{align*}
\sum_{ B^e \ni \, i}\int_{ B^e} \nabla{\psi^e_i} \boldsymbol \cdot \mathcal{G}(\bm{W}_h) \,d B^e\, \approx \, 
	\sum_{ B^e \ni \, i} |B^e| \nabla{\psi^e_i} \boldsymbol \cdot \mathcal{G}(\bm{V}^{e}, \nabla \bm{V}^{e})\,,
\end{align*}
due to the fact that the gradients of the piecewise-linear shape functions $\psi_i^e$ are constant in each primal element of the CFD mesh, $B^e$, whose volume is denoted here by $|B^e|$, and
where $\bm{V}^{e}$ denotes an ``element-value'' of the vector of primitive fluid state variables. FIVER computes $\bm{V}^{e}$ and its gradient $\nabla \bm{V}^{e}$ as follows
\begin{align}
\label{eq:FIVER-Visc-Pop}
	\bm{V}^{e} = \frac{1}{N^e}\sum_{\substack{i \in B^e}}\bm{V}_{i}\,,  \quad \nabla \bm{V}^e = \frac{1}{N^e}\sum_{\substack{i \in B^e}} \nabla{\psi^e_i} \bm{V}_{i}\,,
\end{align}
where $N^e$ denotes the number of grid points connected to the element $B^e$. If $B^e$ is intersected by an impermeable wall $\Omega$, FIVER reconstructs $\bm{V}^{e}$ by populating the velocity and 
temperature at the grid points connected to $B^e$ and occluded by $\Omega$ using constant or linear extrapolation procedures~\cite{lakshminarayan2014embedded, huang2018family}. This leads to the
reconstructed vector $\bm{V}^{e, \textrm{be}}$. Hence, in general, FIVER computes $\bm{V}^e$ and $\nabla \bm{V}^e$ as follows
\begin{equation}
	\bm{V}^{e, {\textrm{be}}} = \frac{1}{N^e}\left(\sum_{\substack{i \in B^e \\ i \textrm{ is not occluded}}}\bm{V}_i + \sum_{\substack{i \in B^e \\ i \textrm{ is occluded}}}
	\bm{V}_i^{\textrm{be}}\right) \quad\textrm{and}\quad \nabla \bm{V}^{e, \textrm{be}} = \sum_{\substack{i \in B^e \\ i \textrm{ is not occluded}}} \nabla {\psi^e_i}\bm{V}_i + 
	\sum_{\substack{i \in B^e \\ i \textrm{ is occluded}}} \nabla{\psi^e_i}\bm{V}_i^{\textrm{be}}.
\end{equation}
It follows that if $\Omega$ is a porous wall, the homogenized approximation of the viscous fluxes at a grid point $i$ inside the porous wall can be computed as follows
\begin{align}
	\sum_{ B^e \ni \, i}\int_{ B^e} \nabla{\psi^e_i} \boldsymbol \cdot \mathcal{G}(\bm{W}_h) \,d B^e = \sum_{ B^e \ni \, i} |B^e| \Big(\alpha \nabla{\psi^e_i} \boldsymbol 
	\cdot \mathcal{G}(\bm{V}^{e}, \nabla \bm{V}^{e}) + (1 - \alpha) \nabla{\psi^e_i} \boldsymbol \cdot \mathcal{G}(\bm{V}^{e, \textrm{be}}, \nabla \bm{V}^{e, \textrm{be}}) \Big).
\end{align}

The body force term (see \Cref{eq: source term}), which acts in the direction normal to the porous wall, is added at each grid point of the embedding CFD mesh
as follows
\begin{equation}
	\bm{S}(\bm{X}) = \begin{pmatrix}
      0 \\
		\eta_f D(\bm{X}) \left( \overline{\frac{\partial \tau_{xy}}{\partial y}} + \overline{\frac{\partial \tau_{xz}}{\partial z}} \right)(\overline u) \, n_{\textrm{w}}\\
      0 \end{pmatrix}\,,
\end{equation}
where $\bm{X}$ denotes position vector of a point in the computational fluid domain $B$, $n_{\textrm{w}}$ is the normal to the porous wall at the closest point $\bm{X}_0 \in \Omega$ to $\bm{X}$,
\begin{equation}
	\overline{u} = \bm{v} \cdot n_{\textrm{w}}\,,
\end{equation}
and $D(\bm{X})$ is the following Gaussian approximation of the Dirac function based on the distance $\|\bm{X} - \bm{X}_0\|_2$ to the porous wall
\begin{equation}
	D(\bm{X}) = \exp{\left[-\frac{1}{2} \left(\frac{(\bm{X} - \bm{X}_0)\cdot n_{\textrm{w}}}{\sigma} \right)^2 \right]}\,
\end{equation} 
where, as before, $\sigma = \eta_f/(2\pi)$.

Finally, it is noted that for the sake of computational efficiency, it suffices in practice to add the above source term only at the grid points located within one layer of elements of the embedding
CFD mesh away from the porous wall.

\section{Pore-resolved numerical simulations}
\label{sec:DNS}

\begin{figure}[!ht]
\centering
\includegraphics[width=0.49\linewidth]{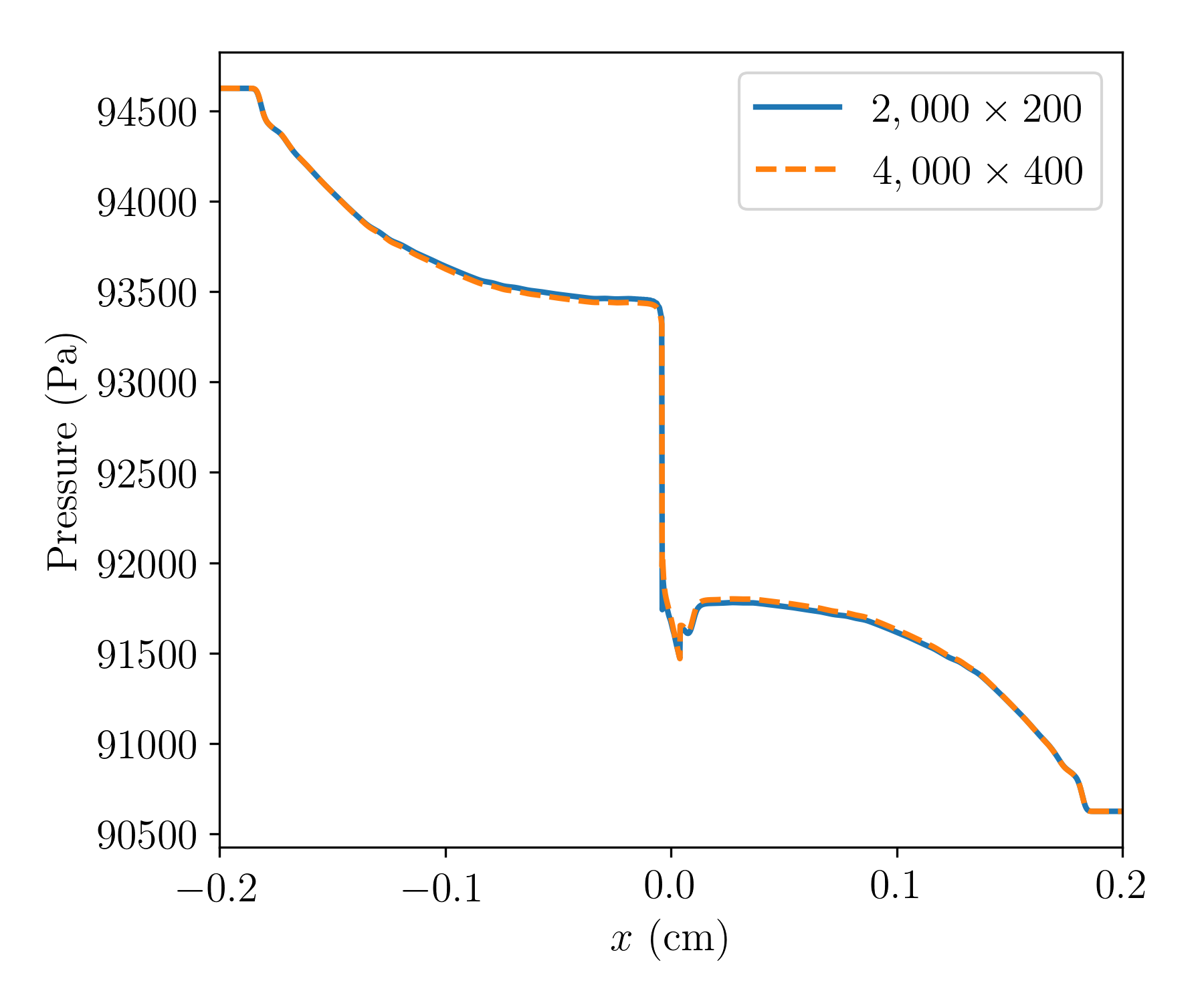}
\includegraphics[width=0.49\linewidth]{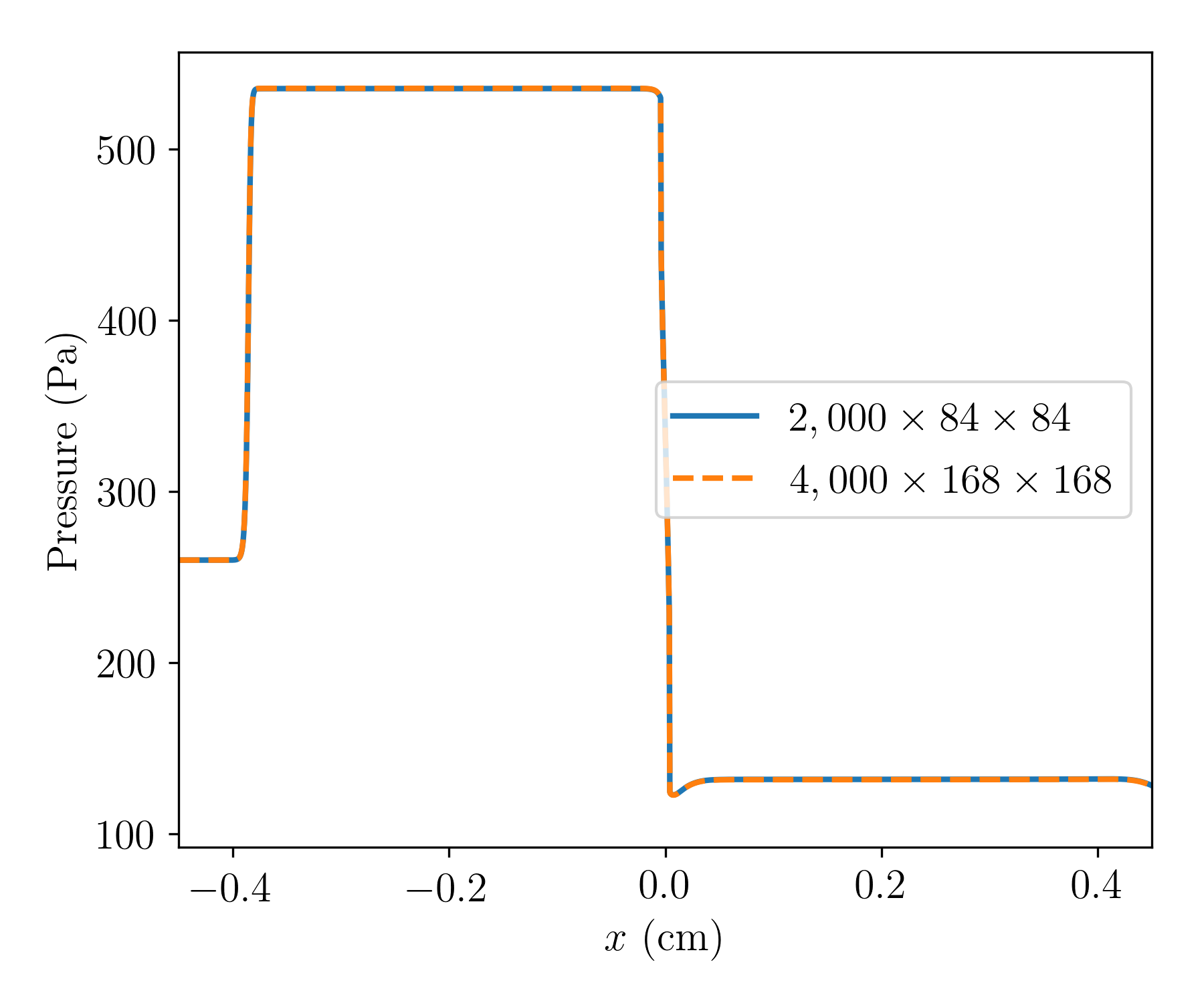}
\caption{Convergence of a pore-resolved simulation: type ``evaluation of a candidate equivalent pore'', $8\% \times 100\%$ pore geometry, initial pressure jump of $4,000 \ \mathrm{Pa}$,
and pressure profile in the central $x$-$y$ plane at $t = 5.3 \ \mathrm{\mu s}$ (left); and type ``shock-porous-wall interaction'', $2/7 \times 2/7$ pore geometry, $V_s=279.4\ \mathrm{m\ s^{-1}}$, 
and pressure profile in the central $x$-$y$ plane at $t = 23.46 \ \mathrm{\mu s}$ (right).}
\label{fig: mesh convergence}
\end{figure}

All pore-resolved numerical simulations discussed in \Cref{sec: app} are performed using the flow solver HAMeRS \cite{wong2017high, wong2016multiresolution, wong2019high} on uniform meshes. 
For each different simulation and pore geometry, the size of the computational domain graphically depicted in \Cref{fig: computational domain} -- or \Cref{fig: Mars initial flow}, as appropriate -- and 
its discretization are reported in \Cref{tab: mesh info}. Specifically, each tabulated discretization is uniform in each direction and verified to be converged by comparing the results it delivers
to their counterparts obtained on a discretization that is twice finer in each direction (for example, see \Cref{fig: mesh convergence}).

\begin{table}[!ht]
\begin{center}
\begin{tabular}{ c|c|c|c } 
\hline
\hline
	Simulation description                      & Pore geometry       & Computational             & Mesh size \\
	                                            &                     & domain (mm)               &           \\
\hline
Evaluation of a candidate equivalent pore (\Cref{sec: exp}) & $2/7 \times 2/7$    & $4 \times 0.5 \times 0.5$  & $800  \times 182 \times 182$  \\
Evaluation of a candidate equivalent pore (\Cref{sec: exp}) & $20\% \times 40\%$  & $4 \times 0.5 \times 0.5$  & $800  \times 200 \times 200$  \\
Evaluation of a candidate equivalent pore (\Cref{sec: exp}) & $10\% \times 80\%$  & $4 \times 0.5 \times 0.5$  & $800  \times 200 \times 200$  \\
Evaluation of a candidate equivalent pore (\Cref{sec: exp}) & $8\% \times 100\%$  & $4 \times 0.5           $  & $1,600 \times 200$ \\
Shock-porous-wall interaction (\Cref{sec:VSA})      & $2/7 \times 2/7$    & $10 \times 0.5 \times 0.5$ & $2,000 \times 84 \times 84$ \\
Shock-porous-wall interaction (\Cref{sec:VSA})      & $8\% \times 100\%$  & $10 \times 0.5$            & $2,000 \times 200$  \\
\hline
\hline
\end{tabular}
\caption {Pore-resolved numerical simulations: computational domains and mesh sizes.}
\label{tab: mesh info}
\end{center}
\end{table}

\end{appendices}

\newpage

\section*{Acknowledgments}
Daniel Z. Huang and Charbel Farhat acknowledge partial support by the Jet Propulsion Laboratory (JPL) under Contract JPL-RSA No. 1590208, and partial support by the
National Aeronautics and Space Administration (NASA) under Early Stage Innovations (ESI) Grant NASA-NNX17AD02G.

\bibliography{references}
\bibliographystyle{plainnat}

\end{document}